\documentclass{aa} % for a referee version
\usepackage{graphicx}
\usepackage{txfonts}
\usepackage{siunitx,subfig,xspace,natbib}
\usepackage{hyperref}
\usepackage{amsmath}%
\usepackage{amsfonts}%
\usepackage{amssymb}%
\usepackage{mathrsfs}
\usepackage{arydshln}
\usepackage{cprotect}
\usepackage{multirow}

\newcommand{\Al}{\element[][26]{Al}\xspace}
\newcommand{\Fe}{\element[][60]{Fe}\xspace}
\newcommand{\degree}{$^{\circ}$}
\newcommand{\Msol}{M\ensuremath{_\odot}\xspace}

\newcommand{\flux}{ph\,cm\ensuremath{^{-2}}\,s\ensuremath{^{-1}}\xspace}

\newcommand{\Msolyr}{M\ensuremath{_\odot}\,yr\ensuremath{^{-1}}\xspace}

\newcommand{\kms}{km\,s$^{-1}$\xspace}
\newcommand{\gray}{$\gamma$-ray\xspace}
\newcommand{\grays}{$\gamma$-rays\xspace}
\usepackage{acronym}
\acrodef{AGB}{asymptotic giant branch}
\acrodef{CGRO}{Compton Gamma Ray Observatory}
\acrodef{COMPTEL}{Imaging Compton Telescope}
\acrodef{INTEGRAL}{International Gamma-Ray Astrophysics Laboratory}
\acrodef{ISM}{interstellar medium}
\acrodef{MHD}{magnetohydrodynamic}
\acrodef{SN}{supernova}
\acrodefplural{SN}[SNe]{supernovae}
\acrodef{GMC}{giant molecular cloud}
\acrodefplural{GMC}[GMCs]{giant molecular clouds}
\acrodef{PDF}{probability distribution function}
\acrodefplural{PDF}[PDFs]{probability distibution functions}
\acrodef{IGM}{intergalactic medium}
\acrodef{ROI}{region of interest}
\acrodefplural{ROI}[ROIs]{regions of interest}
\acrodef{SLUG}{Stochastically Lighting Up Galaxies}
\acrodef{SPI}{Spectrometer on INTEGRAL}
\acrodef{PSYCO}{Population SYnthesis COde}
\acrodef{SFR}{star formation rate}
\acrodef{IMF}{initial mass function}
\acrodef{ECMF}{embedded cluster mass function}
\acrodef{ZAMS}{zero-age main sequence}
\acrodef{MC}{Monte Carlo}

\newcommand{\mrm}[1]{\mathrm{#1}}
\newcommand{\nuc}[2]{$\mrm{^{#2}#1}$}

\usepackage{xcolor}
\definecolor{purple2}{rgb}{0.5, 0.0, 0.5}

\begin{document}

   \title{Galactic Population Synthesis of Radioactive Nucleosynthesis Ejecta}

\author{
    Thomas Siegert \inst{\ref{inst:jmu},\ref{inst:mpe}} \and
	Moritz M. M. Pleintinger \inst{\ref{inst:hnc},\ref{inst:mpe}} \and
    Roland Diehl \inst{\ref{inst:mpe}} \and
    Martin G. H. Krause \inst{\ref{inst:uoh}} \and
    Jochen Greiner \inst{\ref{inst:mpe}} \and
    Christoph Weinberger \inst{\ref{inst:iabg},\ref{inst:mpe}}
}

\institute{
    Institut f\"ur Theoretische Physik und Astrophysik, Universit\"at W\"urzburg, Campus Hubland Nord, Emil-Fischer-Str. 31, 97074 W\"urzburg, Germany
    \label{inst:jmu}
    \and
	Max-Planck-Institut f\"ur extraterrestrische Physik, Giessenbachstra\ss e 1, 85748 Garching b. M\"unchen, Germany
	\label{inst:mpe}
    \and
    Horn \& Company Financial Services GmbH, Kaistraße 20, 40221 Düsseldorf, Germany
    \label{inst:hnc}
    \and
    Centre for Astrophysics Research, University of Hertfordshire, College Lane, Hatfield AL10 9AB, UK
    \label{inst:uoh}
    \and
    Industrieanlagen-Betriebsgesellschaft mbH, Einsteinstr. 20, 85521 Ottobrunn, Germany
    \label{inst:iabg}
}

\date{\textit{Received}: 08.07.2022; \textit{Accepted}: 23.01.2023; \textit{Published}: -.}

\abstract{
Diffuse \gray line emission traces freshly produced radioisotopes in the interstellar gas, providing a unique perspective on the entire Galactic cycle of matter from nucleosynthesis in massive stars to their ejection and mixing in the \ac{ISM}.
We aim at constructing a model of nucleosynthesis ejecta on galactic scale which is specifically tailored to complement the physically most important and empirically accessible features of \gray measurements in the MeV range, in particular for decay \grays such as \nuc{Al}{26}, \nuc{Fe}{60} or \nuc{Ti}{44}.
Based on properties of massive star groups, we developed a \ac{PSYCO} which can instantiate galaxy models quickly and based on many different parameter configurations, such as the \ac{SFR}, density profiles, or stellar evolution models.
As a result, we obtain model maps of nucleosynthesis ejecta in the Galaxy which incorporate the population synthesis calculations of individual massive star groups.
Based on a variety of stellar evolution models, supernova explodabilities, and density distributions, we find that the measured \Al distribution from INTEGRAL/SPI can be explained by a Galaxy-wide population synthesis model with a \ac{SFR} of $4$--$8$\,\Msolyr and a spiral-arm dominated density profile with a scale height of at least 700\,pc.
Our model requires that most massive stars indeed undergo a \ac{SN} explosion.
This corresponds to a \ac{SN} rate in the Milky Way of 1.8--2.8 per century, with quasi-persistent \Al and \Fe masses of $1.2$--$2.4$\,\Msol and 1--6\,\Msol, respectively.
Comparing the simulated morphologies to SPI data suggests that a frequent merging of superbubbles may take place in the Galaxy, and that an unknown but strong foreground emission at 1.8\,MeV could be present.
}


\maketitle
%
%-------------------------------------------------------------------

\section{Introduction}\label{sec:intro}

The presence of radioisotopes in interstellar gas shows that the Milky Way is continuously evolving in its composition.
This aspect of Galactic evolution proceeds in a cycle of star formation, nucleosynthesis feedback, and large-scale mixing.
Radioisotopes trace this entire process through their \gray imprints, i.e. the production, ejection, and distribution of radionuclides.
Thus, the investigation of radioactivity in the Galaxy provides an astrophysical key to the interlinkage of these fundamental processes.

The currently most frequently and thoroughly studied \gray tracers of nucleosynthesis feedback are the emission lines of \Al and \Fe ejecta from stellar winds and \acp{SN}.
Their half-life times of 0.7\,Myr \citep{Norris:1983al} and 2.6\,Myr \citep{Rugel:2009fe}, respectively, are comparable to the dynamic timescales of superbubble structures around stellar groups \citep{deAvillez:2005hd, Keller:2014sb, Keller:2016sb}.
The nuclear decay radiation carries a direct signature of the physical connection between their production in massive stars and their distribution in the \ac{ISM}.
Spatial mapping of the interstellar \Al emission at 1809\,keV \citep[e.g.,][]{1995A&A...298..445D,Oberlack:1996ag, Pluschke:2001voa, Bouchet:2015aa} and detailed \gray line spectroscopy \citep[e.g.,][]{Diehl:2006aa, Diehl:2010bc, Kretschmer:2013aa, Siegert:2017aa, Krause:2018aa} provided observational insights into the relation between nucleosynthesis ejecta and the dynamics of massive star groups.
Together with the detection of \Fe emission lines from the \ac{ISM} at 1173\,keV and 1332\,keV \citep{Wang:2007aa, Wang:2020fe}, this represents an indispensable astrophysical effort to understand the feedback cycle underlying Galactic chemical enrichment.

In order to follow and understand the complex dynamics, astrophysical models and simulations have to be utilised.
Because directly mapping the diffuse emission of the 1.8\,MeV line is difficult and image reconstructions can come with considerable bias, given the photon-counting nature of the measurements, empirical and descriptive models of \Al in the Galaxy have been invoked in the past \citep[e.g.,][]{Prantzos:1993mn, Prantzos:1995gd, Knodlseder:1996wh, Lentz:1999ui, Sturner:2001yu, Drimmel:2002il, Alexis:2014ux}.
Their scientific interpretations rely on comparisons between the measured \Al emission and multi-wavelength `tracers' or geometric (smooth) emission morphologies \citep[e.g.,][]{Hartmann:1994eo, Prantzos:1995gd, Diehl:1997op, Knodlseder:1999we, Diehl:2004ke, Kretschmer:2013aa}.
Such heuristic approaches can come with two major downsides:
On the one hand, comparisons with tracer maps at other wavelengths include many astrophysical assumptions, such as the ionisation of massive stars that produce \nuc{Al}{26} that finally lead to free-free emission \citep{Knoedlseder1999_freefree1.8MeV}, which might put these comparisons on shaky grounds.
On the other hand, descriptive models, such as doubly-exponential disks, contain hardly any astrophysical input, because only extents are determined.
Heuristic comparisons therefore offer only limited potential for astrophysical interpretation.
These earlier studies focussed more on the general description of the overall \gray morphology because it was (and still is) unclear where exactly the emission originates.
\citet{Prantzos1996_Al26_theory} first discussed the different contributions to the Galactic \nuc{Al}{26} signal, distinguishing between massive stars, AGB stars, classical novae, and cosmic-ray spallation.
The ratio of \nuc{Al}{26}-production from these earlier estimates is $1:2:3$ for AGB, massive stars and nova contributions in units of $\mathrm{M_{\odot}\,Myr^{-1}}$ with a negligible cosmogenic production.
From comparisons to COMPTEL data, \citet{Knoedlseder1999_freefree1.8MeV} finds that the nova and AGB contribution may both only be up to $0.2\,\mathrm{M_{\odot}\,Myr^{-1}}$, whereas between 80--90\,\% originate from massive stars.
Recent modelling of classical novae by \citet{Bennett2013_nova26Al}, taking into account the resonance interaction $\mathrm{^{25}Al(p,\gamma)^{26}Si}$ suggest a nova contribution to the total \nuc{Al}{26} mass in the Milky Way of up to 30\,\%.
There is no resolved measurement of the \nuc{Al}{26}-profile around Wolf-Rayet stars \citep{Mowlavi:2006gv}, for example, and only a few measurements of the distribution inside superbubbles \citep{Plueschke01PhD,Diehl:2003kk,Krause:2018aa} which can gauge the stellar evolution models.
Owing to the instruments' capabilities with a typical angular resolution of 3\degree, this is understandable.
However, with more than 17 years of data, the descriptive parameters, such as the total \Al line flux or the scale height of the disk \citep[e.g.,][]{Pleintinger:2019tq} are now precisely determined, which allows us to ask more fundamental questions.

In this study, we attempt to shift the focus from interpretations of descriptive parameters to astrophysical input parameters to describe the \Al sky.
Modelling an entire Galaxy for comparisons to \gray data appears intractable because hydrodynamics simulations suggest interdependencies that could hardly be modelled:
First in-depth simulations regarding the Galaxy-wide distribution of nucleosynthesis ejecta have been performed by \citet{Fujimoto:2018aa, Fujimoto:2020al} and \citet{Rodgers-Lee:2019al}.
These simulations start from the description of basic physical conditions, such as the gravitational potential of a galaxy and its temperature and density profile to solve the hydrodynamics equations, and follow the spread of freshly synthesised nuclei in a simulated galaxy.
Comparing such simulations directly to \gray data yields insights into the astrophysical 3D modalities and dynamics underlying the measured radioactivity in the Milky Way \citep{Pleintinger:2019tq}.
A scientific exploitation of such simulations, however, faces major limitations:
On the one hand, only a few realisations can be created because they are computationally expensive.
Due to their particular characteristics (generalising the Galaxy, rather than accounting for the location of the Sun and nearby spiral arms, for example), these are only comparable to the Milky Way in a limited extent.
On the other hand, current \gray instruments have poor sensitivities, so that the level of detail of hydrodynamic simulations cannot be covered by observations.

In this paper, we present an alternative approach to modelling the radioactive Galaxy that is specifically adapted to the empirical basis of \gray measurements.
The essential scientific requirements here are that it can be repeated quickly and in many different parameter configurations and that, at the same time, the astrophysically most important and observationally accessible features can be addressed.
For this purpose, we developed the \acf{PSYCO}, the general structure of which we outlined in Sect.\,\ref{sec:model_params}.
It is designed as bottom-up model of nucleosynthesis ejecta in the Galaxy, and based on population synthesis calculations of massive star groups, which will be described in Sect.\,\ref{sec:popsyn}.
Our simulation results are shown in Sect.\,\ref{sec:performance} and quantitative comparisons to the entire Galaxy are outlined in Sect.\,\ref{sec:comparison}, in particular summarising the astrophysical parameters of interest that can be obtained with \gray observations from INTEGRAL/SPI \citep{Winkler:2003aa,Vedrenne:2003aa}.
We discussion our findings in Sect.\,\ref{sec:discussion} and summarise in Sect.\,\ref{sec:outlook}.

\section{Model Structure and Input Parameters}\label{sec:model_params}

 \begin{figure*}[t!]
	\centering
	\includegraphics[width=17cm, trim=9cm 4cm 9cm 4.5cm, clip]{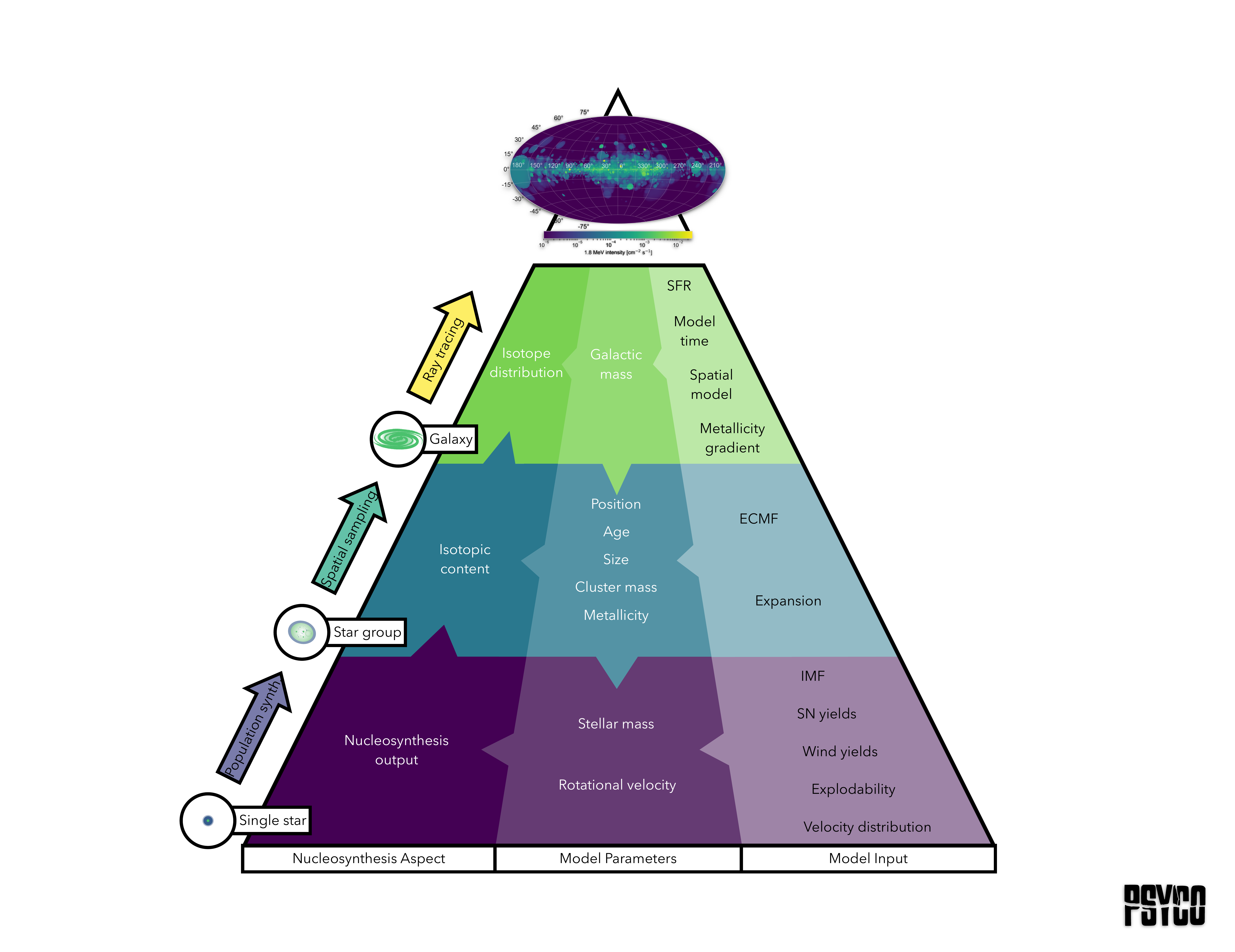}
	\caption{Structure of the \ac{PSYCO} model. Using the model input on the right, model parameters are accumulated top-down. The nucleosynthesis aspect on the left is finally built bottom-up to construct all-sky \gray maps.}
	\label{fig:psyco_structure}
\end{figure*}

The overall model structure of \ac{PSYCO} is shown in Fig.\,\ref{fig:psyco_structure} \citep{Pleintinger2020_PhD}.
While the input parameters are determined top-down from the galactic level to single star properties, the nucleosynthesis aspect is subsequently modelled bottom-up with population synthesis calculations of massive star groups (Sect.\,\ref{sec:popsyn}).

\subsection{Galactic Scale}\label{sec:gal_scale}
\subsubsection{Stellar mass and timescales}\label{sec:stellar_mass}
The overall \ac{SFR} of the Milky Way remains debated, as it can be determined in many different ways and their results are almost as varied.
For example, interpretations of HII regions obtain a value of $(1.3 \pm 0.2)$\,\Msolyr \citep{Murray:2010sf}, of infrared measurements a value of 2.7\,\Msolyr \citep{Misiriotis:2006sf}, of \grays a value of 4\,\Msolyr \citep{Diehl:2006aa}, or of HI gas a value of 8.25\,\Msolyr \citep{Kennicutt:2012hg}.
Meta-analyses seem to settle on a value between 1.5--2.3\,\Msolyr \citep{Chomiuk:2011sf, Licquia:2015sf}.
Likewise, a hydrodynamics-based approach by \citet{Rodgers-Lee:2019al} finds a range of 1.5--4.5\,\Msolyr.
Because the \ac{SFR} is an important model input variable, we use it as a free parameter.

The necessary model time is determined by the time scale of the longest radioactive decay (here \Fe).
We start from an `empty' galaxy which we gradually fill with mass in stars and radioactive ejecta.
After the initial rise, owing to the constant \ac{SFR}, a quasi-constant production rate balances a quasi-constant decay rate in the simulated galaxy.
This leads to specific \gray line luminosities which can be compared to data.
We exploit the fact that the galactic amount of species $i$ with lifetime $\tau_i$ will approach this equilibrium between production and decay after some time which is related to the supernova rate which, in turn, is related to the stellar models used.
In total, after $T_{\text{tot}} \approx 50$\,Myr the galactic amount of the longer-lived \Fe has decoupled from the initial conditions and reached a quasi-equilibrium.
This defines the total model time and accordingly the total mass $M_{\text{gal}} = \text{SFR} \times T_{\text{tot}}$, which is processed into stars during that time.
Treating the constant \ac{SFR} as free parameter sets the main physical boundary on the stellar mass formed in the model.

\subsubsection{Spatial Characteristics}\label{subsubsec:spatial_characteristics}
In order to ensure fast calculation, we use the following assumptions about the overall morphology and the metallicity gradient at the Galactic level:
During the expected visibility of an individual star group of $\sim25$\,Myr as defined by the decay of the \Fe ejected from the last supernova (see Sect.\,\ref{sec:star_group}), the Milky Way rotates by about 38\degree.
This is short enough that the position of the observer as well as gas and stars can be assumed as approximately co-rotating with respect to the spiral arms.
Thus, a static galactic morphology divided in a radial and a vertical component is adopted.
In the Galactic coordinate frame the Sun is positioned 8.5\,kpc from the Galactic centre and 10\,pc above the plane \citep{Reid:2016mw, Siegert:2019vp}.
We note that the Galactic centre distance may be uncertain by up to 8\,\%, for which the resulting flux values and in turn luminosity, mass, and SFR may be uncertain by 14\,\%.
The vertical star formation density is chosen to follow an exponential distribution $\rho(z; z_0) = z_0^{-1}\exp(-|z|/z_0)$, with the scale height $z_0$ parametrising the Galactic disk thickness.
The radial star formation density of the Galaxy can be approximated by a truncated Gaussian \citep{Yusifov:2004mw},
\begin{equation}\label{eq:spiral_radial_distribution}
\rho(R; R_{\mu},\sigma) = \begin{cases}
\frac{1}{\sqrt{2\pi}\sigma}\ \exp{\left[-\frac{(R - R_{\mu})^2}{2\sigma^2}\right]}, &\text{if}\ R \leq 20\,\mrm{kpc}\\
0, &\text{else,}
\end{cases}
\end{equation}
with a maximum at radius $R_{\mu}$ and width $\sigma$.
Alternatively, an exponential profile can be chosen to obtain more centrally concentrated radial morphologies.
The radial distribution is convolved with a 2D structure of four logarithmic spirals, approximating the expected spiral arm structure of the Milky Way.
Each spiral centroid is defined by its rotation angle \citep{Wainscoat:1992mw},
\begin{equation}\label{eq:spiral_arm_centroid}
\theta(R) = k\ \ln\left(\frac{R}{R_0}\right) + \theta_0,
\end{equation}
along the radial variable $R$, with inner radius $R_0$, offset angle $\theta_0$, and widening $k$.
In order to match the observed Milky Way spiral structure \citep{Cordes:2002mw, Vallee:2008mw} the values in Tab.~\ref{tab:spirals} are adopted.
Around the spiral centroids, a Gaussian-shaped spread for star formation is chosen.
In this work, it is generally assumed that the spatial distribution of star groups follows the Galactic-wide density distribution \citep{Heitsch:2008ts, Micic:2013ts, Gong:2015ts, Krause:2018aa}.

\begin{table}[b]
	\caption{Spiral centroid parameters (cf.~Eq.~(\ref{eq:spiral_arm_centroid})) following \citet{Faucher:2006mw}.}             % title of Table
	\label{tab:spirals}      % is used to refer this table in the text
	\centering                          % used for centering table
	\begin{tabular}{l c c c}        % centered columns (4 columns)
		\hline\hline                 % inserts double horizontal lines
		Spiral arm & $k$ [rad] & $R_0$ [kpc] & $\theta_0$ [rad] \\
		\hline
		Norma & $4.25$ & $3.48$ & $3.141$ \\
		Scutum-Centaurus & $4.89$ & $4.90$ & $2.525$ \\
		Sagittarius-Carina & $4.25$ & $3.48$ & $0.000$ \\
		Perseus/Local & $4.89$ & $4.90$ & $5.666$ \\
		\hline                                   %inserts single line
	\end{tabular}
\end{table}

\begin{figure}[!ht]
	\centering
	\includegraphics[width=\hsize,trim=3cm 2cm 1.8cm 2.6cm, clip]{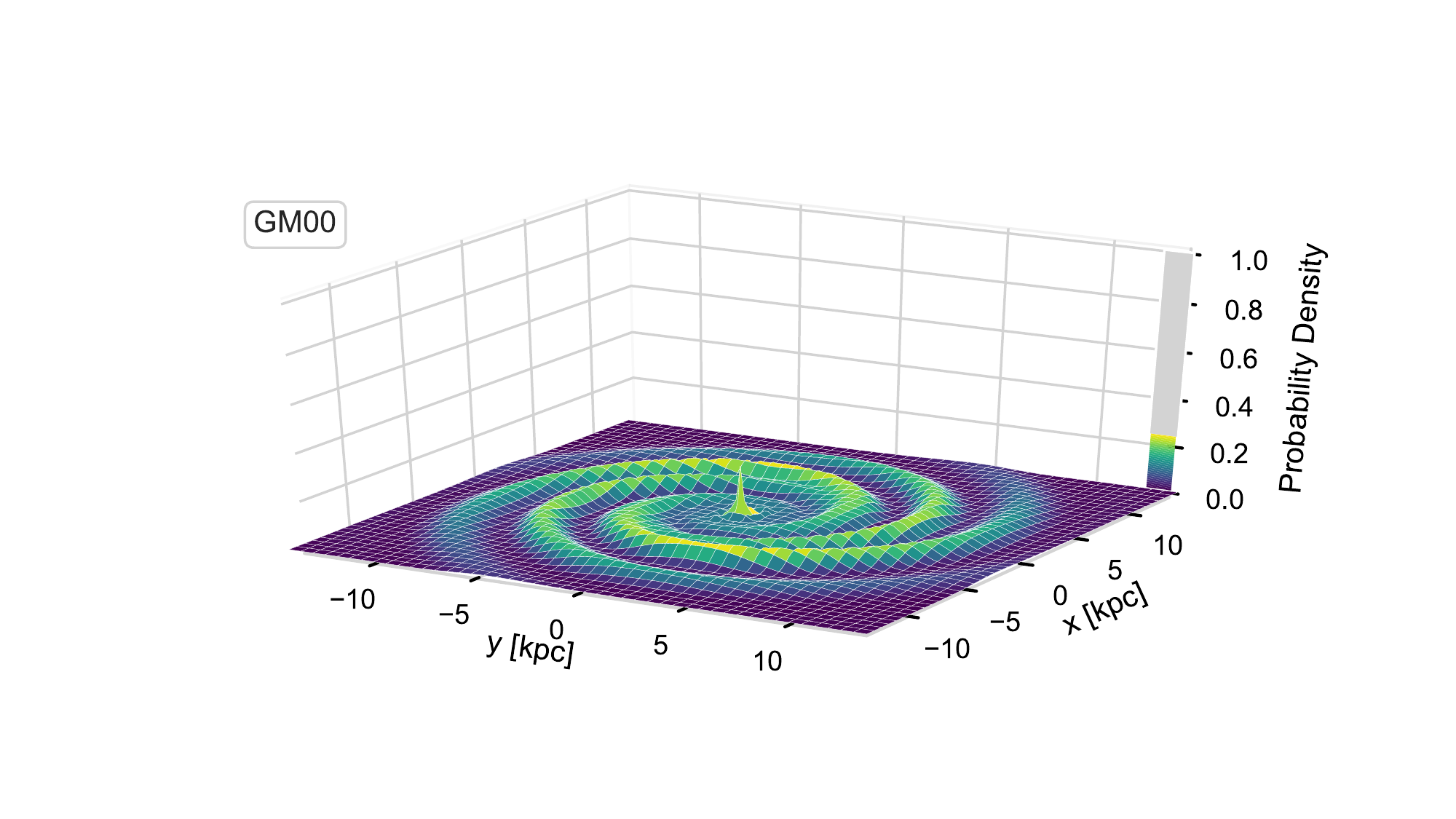}
	\includegraphics[width=\hsize,trim=3cm 2cm 1.8cm 2.6cm, clip]{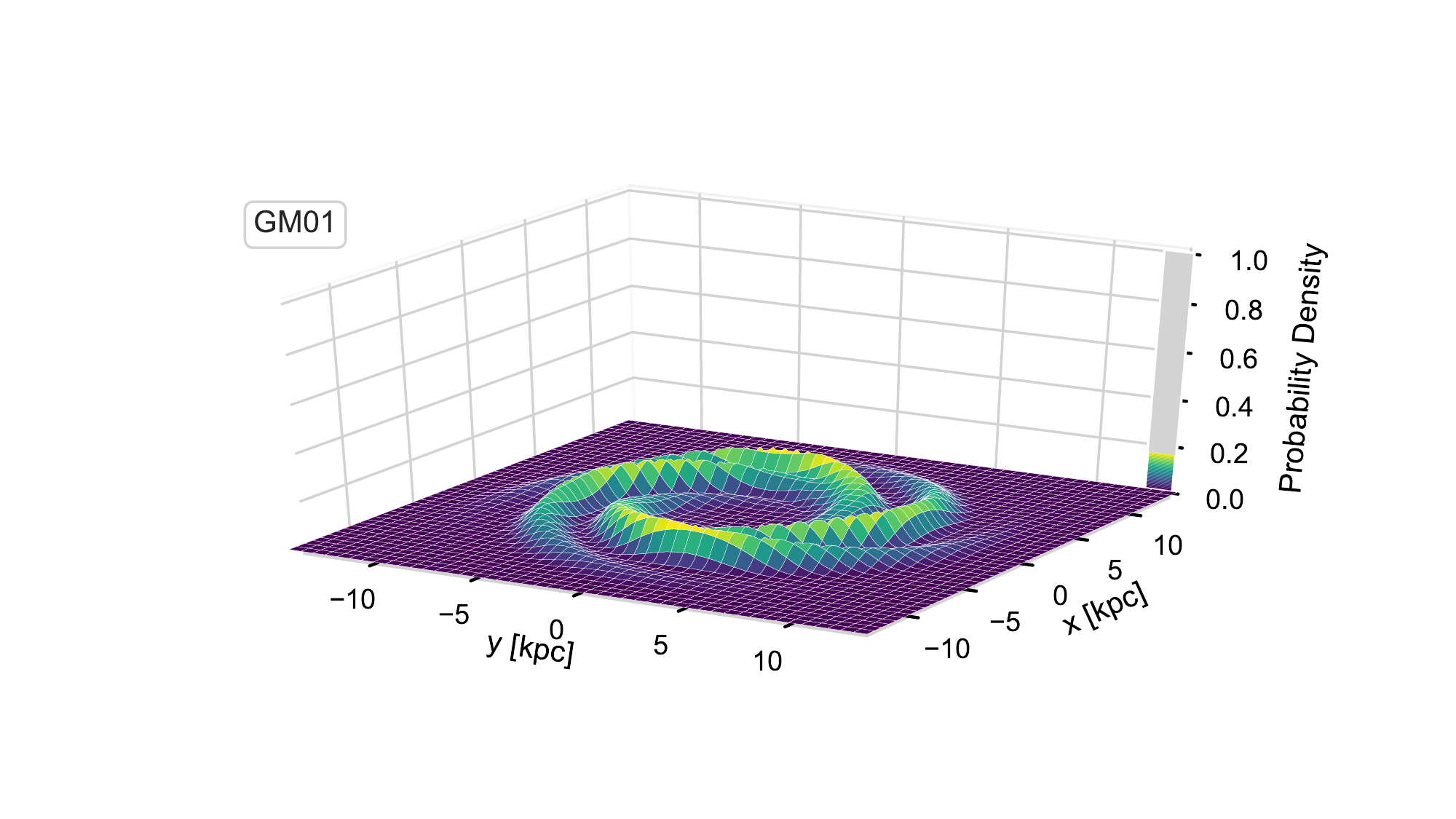}
	\includegraphics[width=\hsize,trim=3cm 2cm 1.8cm 2.6cm, clip]{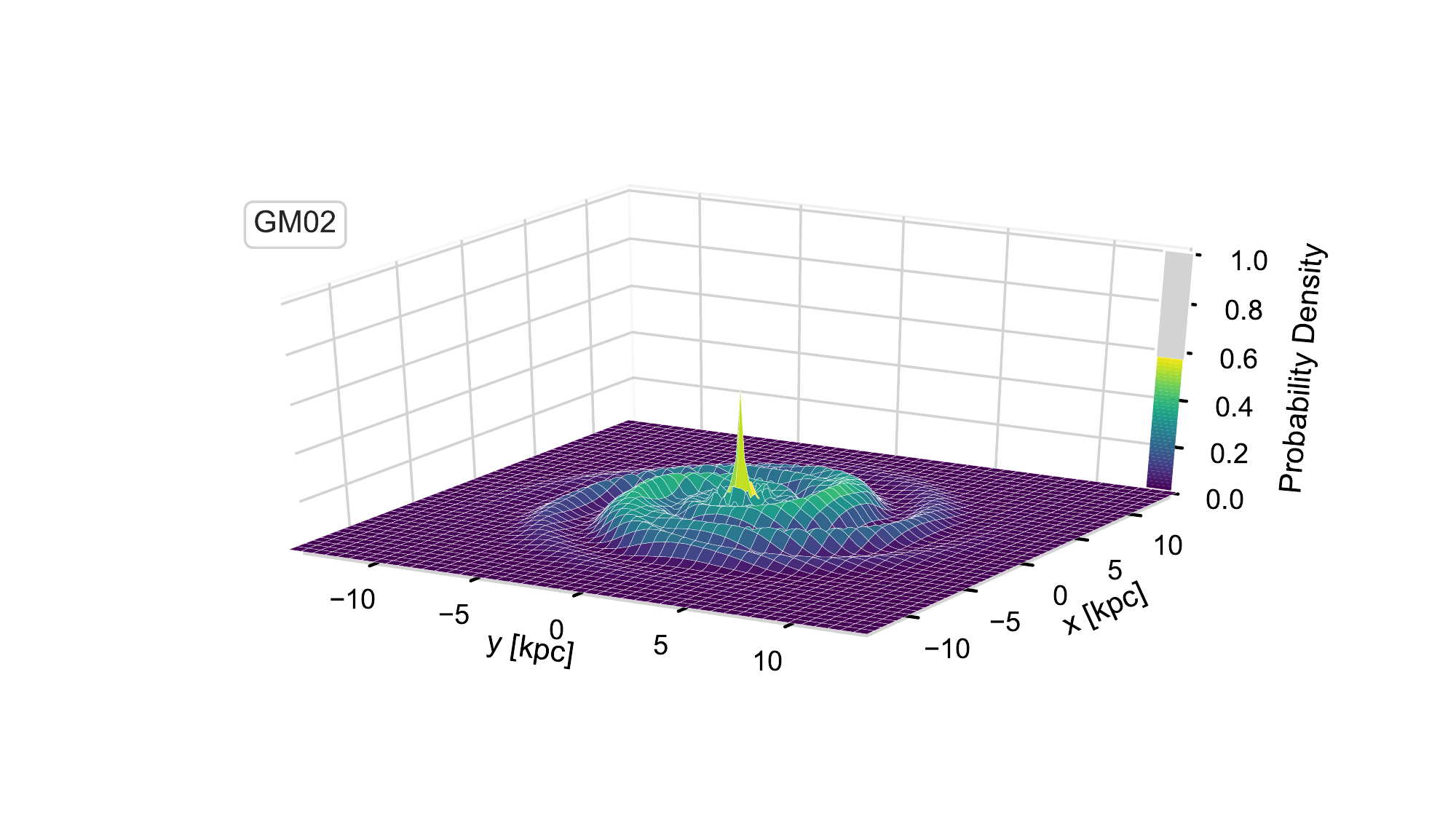}
	\includegraphics[width=\hsize,trim=3cm 2cm 1.8cm 2.6cm, clip]{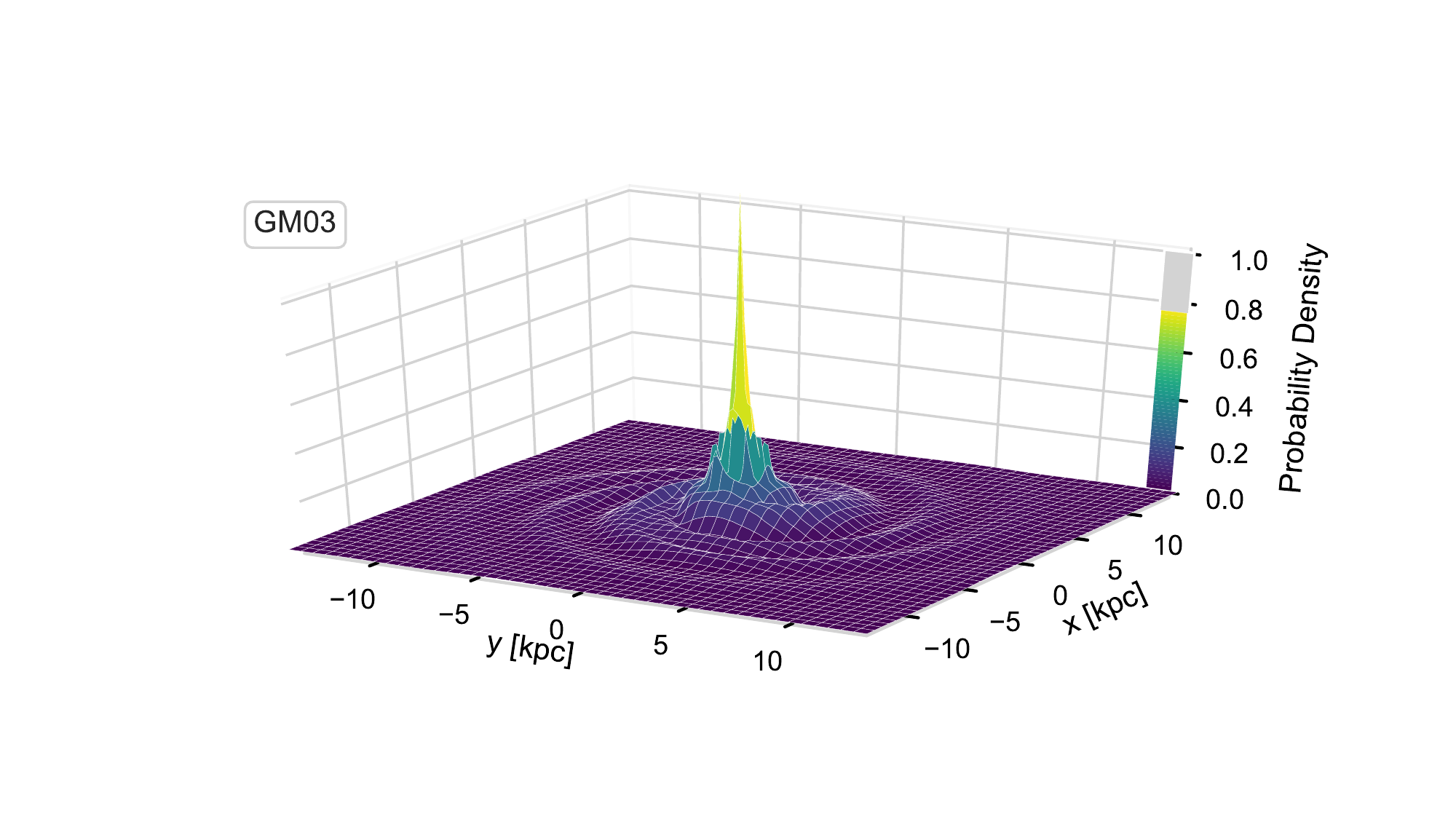}
	\includegraphics[width=\hsize,trim=3cm 2cm 1.8cm 2.6cm, clip]{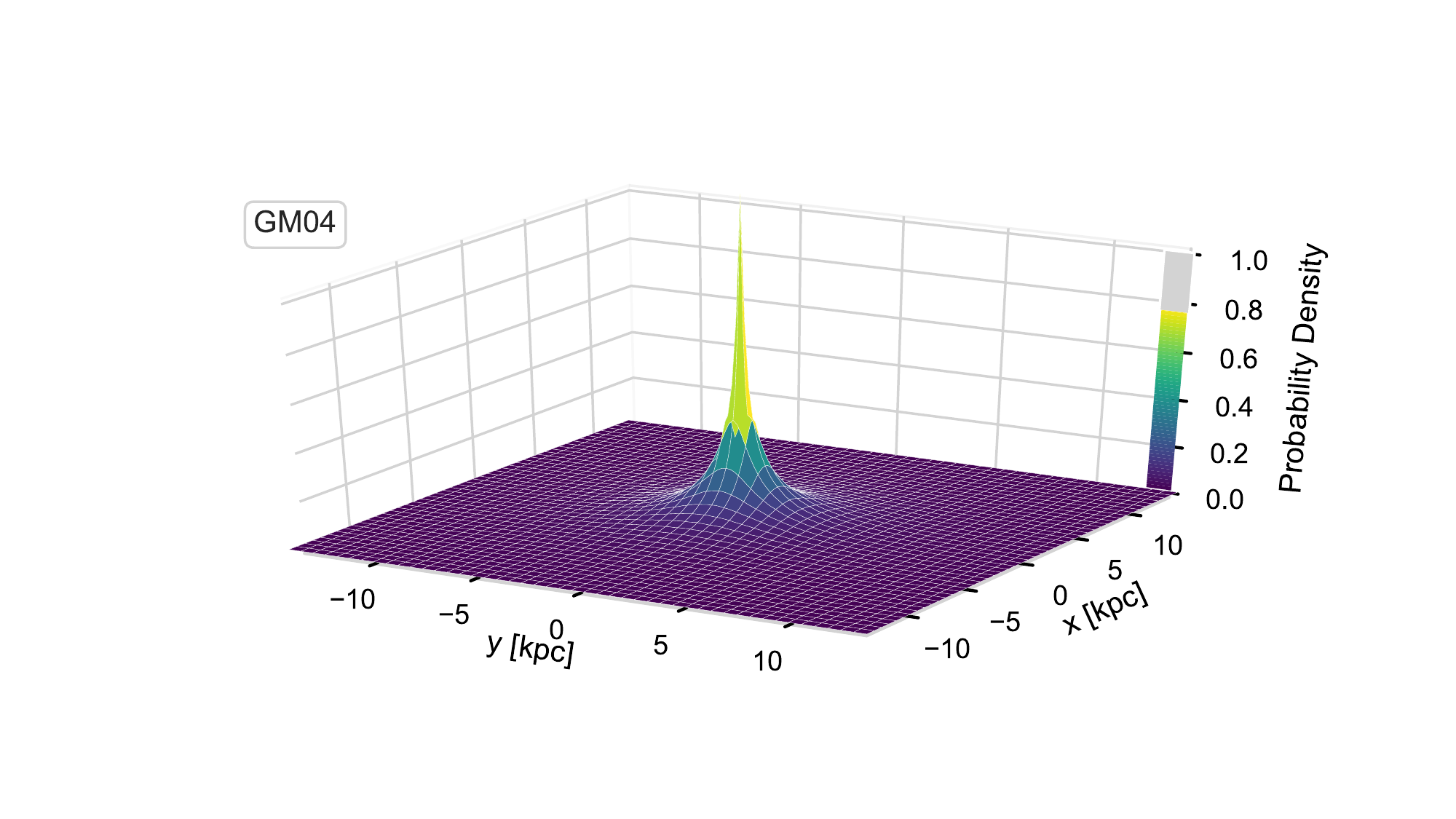}
	\caption{Radial probability density distributions for star formation. The morphologies GM00--GM04 are ordered from top to bottom by decreasing dominance of the Galactic spiral structure. GM04 is a purely exponential disk.}
	\label{fig:radial_distribution}
\end{figure}

Fig.\,\ref{fig:radial_distribution} depicts five different galactic morphologies that are implemented in the model.
From pronounced spiral structures to a smooth central peak, their nomenclature follows GM00--GM04, with GM00--GM02 using a Gaussian and GM03 and GM04 using an exponential density profile in radius.
In the latter case, a scale radius of 5.5\,kpc is adopted.
In particular, GM00 describes a balance of a central density peak (for example from the central molecular zone) with the spiral arms.
GM01 ignores the central peak, as found in \citet{Kretschmer:2013aa} and modelled by \citet{Krause:2015aa}.
GM02 enhances the peak central density by placing $R_{\mu}$ of the Gaussian density closer to the center.
GM03 describes the distribution of pulsars \citep{Faucher:2006mw}, and shows a large exponentially peaked maximum in the center.
Finally, GM04 represents an exponential radial profile, being completely agnostic about spiral features.
The latter is typically used for inferences with \gray telescopes to determine the scale radius and height of the Galaxy in radioactivities.
GM03 and GM04 could also mimic recent, several Myr year old, star burst activity near the Galactic centre as suggested from interpretations of the Fermi bubbles \citep[e.g.,][]{Crocker2012_FermiBubbles_outflows}.
We model the metallicity gradient in the Milky Way because it is measured to generally decrease radially \citep[e.g.,][]{Cheng:2012mt}.
For the modelling of nucleosynthesis ejecta, this is an important factor because it determines the amount of seed nuclei and the opacity of stellar gas.
To include this effect on nuclear yields and stellar wind strength, the Galactic metallicity gradients for different heights are included in the model following the measurements of \citet{Cheng:2012mt}.
For different heights above the plane it can be approximated linearly with the parameters found in Tab.\,\ref{tab:metallicity_gradient}.
\begin{table}
	\caption{Linear parameters for modelling the radial metallicity gradient in the Milky Way according to different heights above the Galactic plane \citep{Cheng:2012mt}. The Intersect is in units of the metallicity, $\mrm{[Fe/H]}$, and the Slope in units of $\mrm{[Fe/H]\,kpc^{-1}}$.}             % title of Table
	\label{tab:metallicity_gradient}      % is used to refer this table in the text
	\centering                          % used for centering table
	\begin{tabular}{l c c}        % centered columns (4 columns)
		\hline\hline                 % inserts double horizontal lines
		Height [kpc] & Slope & Intersect \\
		\hline
		$\geq 1.00$& $-0.0028$& $-0.5$ \\
		0.50--1.00 & $-0.013$ & $-0.3$ \\
		0.25--0.50 & $-0.55$ & $0$\\
		0.15--0.25 & $-0.36$ & $0$\\
	\end{tabular}
\end{table}

\subsection{Stellar groups}
\subsubsection{Distribution function of star formation events}
Stars form in more extended groups as well as more concentrated clusters.
Our model does not require information about the spatial distribution on cluster scale, so that we define `star formation events'.
Following \citet{Krause:2015aa} we assume a single distribution function for all kinds of star forming events, $\xi_{\rm EC}$ with mass of the embedded cluster, $M_{\text{EC}}$, which is empirically described by a power law
\begin{equation} \label{eq:ecmf}
\xi_{\text{EC}}(M_{\text{EC}}) = \frac{dN_{\text{EC}}}{dM_{\text{EC}}} \propto M_{\text{EC}}^{-\alpha_{\text{EC}}}\mrm{,}
\end{equation}
with $\alpha_{\text{EC}} = 2$ in the Milky Way \citep{Lada:2003il, Kroupa:2013bw, Krumholz:2019sg}.
This applies to star formation events between $5 \leq M_{\text{EC}}/\text{\,\Msol} \leq 10^7$ \citep{PflammAltenburg:2013ib, Yan:2017vj}.
In order to approximate the physical properties of cluster formation, this relation is implemented in the model as probability distribution to describe the stochastic transfer of gas mass into star groups over the course of $T_{\text{tot}} = 50$\,Myr.

\subsubsection{Initial Mass Function}\label{sec:IMF}
Inside a star forming cluster, the number of stars $dN_*$ that form in the mass interval between $M$ and $M + dM$ is empirically described by the stellar \ac{IMF} of the general form
\begin{equation} \label{eq:imf_salpeter}
\xi_{\text{S55}}(M_*) = \frac{dN_*}{dM} =  kM_*^{-\alpha},
\end{equation}
with power-law index $\alpha$ for stellar mass $M_*$ in units of M$_{\odot}$ and a normalisation constant $k$.
The \ac{IMF} itself is not directly measurable.
Thus its determination relies on a variety of basic assumptions and it is subject to observational uncertainties and biases \citep{Kroupa:2013bw}.
Typically used \ac{IMF}s are from \citet[][S55]{Salpeter:1955aa}, \citet[][K01]{Kroupa2001_IMF}, and \citet[][C03]{Chabrier:2003ki}, which we include in our model.
Details about the functional forms on these \ac{IMF}s are found in Appendix\,\ref{sec:appendix_IMF}.

These three variations of the \ac{IMF} are implemented to describe the statistical manifestation of the physical star formation process in each star group.
In all cases, we limit the IMF to masses below 150\,\Msol since observationally higher stellar masses are, so far, unknown \citep{Weidner:2004ms, Oey:2005im, MaizApellaniz:2008im}.
We note that the yield models (Sect.\,\ref{sec:nucsys}) typically only calculate up to 120\,\Msol, so that extrapolations to higher masses are required, which might be unphysical.
Given that the high stellar mass IMF index $x_3$ for stars above $1\,M_{\odot}$ may also be uncertain \citep[$x_3 \approx 3 \pm 1$;][]{Kroupa2021_IMF}, but the fraction of stars above $100\,M_{\odot}$ is at most 0.05\,\%, we consider the estimates robust against changes in the IMF above $100\,M_{\odot}$.
This mass limit, $M_{\text{max}}$, seems to vary with the total cluster mass $M_{\text{EC}}$ \citep{Pflamm:2007ms, Kroupa:2013bw}:
This relation can be approximated by
\begin{equation} \label{eq:maximum_stellar_mass}
\log(M_{\text{max}}) = 2.56 \log(M_{\text{EC}}) \left\{3.821^{\lambda} + \left[\log(M_{\text{EC}})\right]^{\lambda}\right\}^{-1/\lambda} - 0.38,
\end{equation}
with $\lambda = 9.17$.
The upper-mass limit is applied for each star group together with the lower-mass limit for deuterium burning of $0.012$\,\Msol \citep{Luhman:2012lm}.

\subsubsection{Superbubbles}
Radiative, thermal, and kinetic feedback mechanisms from individual stars in associations shape the surrounding \ac{ISM}.
This cumulative feedback produces superbubble structures around each stellar group.
These structures are approximated as isotropically and spherically expanding bubbles \citep{Castor:1975ib,Krause:2014aa}.
The crossing-time of stellar ejecta in such superbubbles is typically $\sim1$\,Myr \citep{Lada:2003il}, which corresponds to the lifetime of \Al.
Due to this coincidence, the distribution of \Al is expected to follow the general dynamics of superbubbles \citep{Krause:2015aa}.
Therefore, the size scale of superbubbles is adopted as a basis for the spatial modelling of nucleosynthesis ejecta in our model.

Hydrodynamic \ac{ISM} simulations by \citet{deAvillez:2005hd}, for example, show that after a similar timespan of $\sim1$\,Myr the hot interior of a superbubble is homogenised.
Thus, we model nucleosynthesis ejecta as homogeneously filled and expanding spheres with radius \citep{Weaver:1977sb, Kavanagh:2020sb}
\begin{equation}\label{eq:bubble_expansion}
R_{\text{bubble}}(t) = x L_{\text{W}}^{1/5} t^{3/5},
\end{equation}
as a function of the mechanical luminosity $L_{\text{W}}$ and time $t$.
The mechanical luminosity was modelled in population synthesis calculations \citep[e.g.,][]{Pluschke:2001ut} with a canonical value of $L_{\text{W}} = 10^{38}$\,erg\,s$^{-1}$, which we adopt in this work.
The free parameter $x$ in Eq.\,(\ref{eq:bubble_expansion}) relates the ISM density, $\rho_0$, and the constant for radial growth over time, $\alpha$ by $x = \alpha \rho_0^{-1/5}$.
Literature values for $\alpha$ range between $0.51$ for enhanced cooling in mixing regions \citep{Krause:2014sc} (see also \citep{Fierlinger2012_ISMfeedback,Fierlinger2014_PhDthesis,Krause:2013aa,Krause:2015aa}) to $0.76$ for the self-similar, analytical, adiabatic expanding solution \citep{Weaver:1977sb}.
We use a constant $x$ of $4 \times 10^3\,\mathrm{kg^{-1/5}\,m^{3/5}}$, which would correspond to a particle density around $100\,\mathrm{cm^{-3}}$ for the self-similar solution and a density around $20\,\mathrm{cm^{-3}}$ for enhanced cooling.
This sets the boundary conditions for the temporal evolution of the individual superbubbles.
Relevant bubbles sizes, i.e. when \Al and \Fe can still be found inside before decay, are therefore on the order of a few 100\,pc.
We discuss the possible biases in resulting fluxes and morphologies in Sect.\,\ref{sec:caveats}, and proceed with the assumptions presented above.

We assume cluster formation to be a stochastic process that occurs with a constant rate over time.
Spatially, it is preferentially triggered, when gas is swept up by the gravitational potential of a spiral arm.
Each newly formed stellar cluster is assigned a 3D position according to a galactic morphology as shown in Fig.\,\ref{fig:radial_distribution}.
The 3D position then implies the metallicty as calculated from the Galactic metallicity gradient (Sect.\,\ref{subsubsec:spatial_characteristics}).
In the model, the fundamental parameters of total mass $M_{\rm EC}$, age $T_{\rm bubble}$, position $(x,y,z)_{\rm bubble}$, and metallicity $Z_{\rm bubble}$ are assigned to each individual star group.

\subsection{Stellar parameters}\label{sec:stellar_scale}
\subsubsection{Stellar rotation}\label{subsubsec:stellar_rotation}
Stellar rotation creates additional advection, turbulent diffusion, and enhances convective regions inside the star \citep{Endal:1978sr, Heger:2000se}, which increases mixing and transport of material inside the star.
Stellar winds are amplified and also occur earlier due to the rotation.
The wind phase as well as the entire lifetime of the stars is also extended.
Stellar evolution models suggest that these effects have a significant impact on nucleosynthesis processes inside the stars as well as on their ejection yields \citep[e.g.,][]{Limongi:2018gh, Prantzos:2019ro, Banerjee:2019ro, Choplin:2020ro}.
We implement stellar rotation for our galactic nucleosynthesis model by the following considerations:

For each star that forms in the model, an additional step is included in the population synthesis process to randomly sample a rotation velocity according to measured distributions.
Stellar rotation properties have been catalogued observationally for each spectral class by \citet{Glebocki:2000vr,Glebocki:2005vr}.
To include this information in population synthesis calculations, the observed rotation velocities are weighted with the average inclination angle of the stars.
The resulting distributions of rotation velocities are then fitted for each spectral class individually by a Gaussian on top of a linear tail.
In this context, the most relevant classes are the massive O- and B-type stars.
They show a maximum at $100$\,\kms with a width of $60$\,\kms (O) and $0$\,\kms and width $180$\,\kms (B), respectively.

\subsubsection{Explodability}\label{subsubsec:explodability}
\begin{figure}[!ht]
	\centering
	\includegraphics[width=\columnwidth,trim=1.5cm 3.5cm 3.5cm 4cm, clip=true]{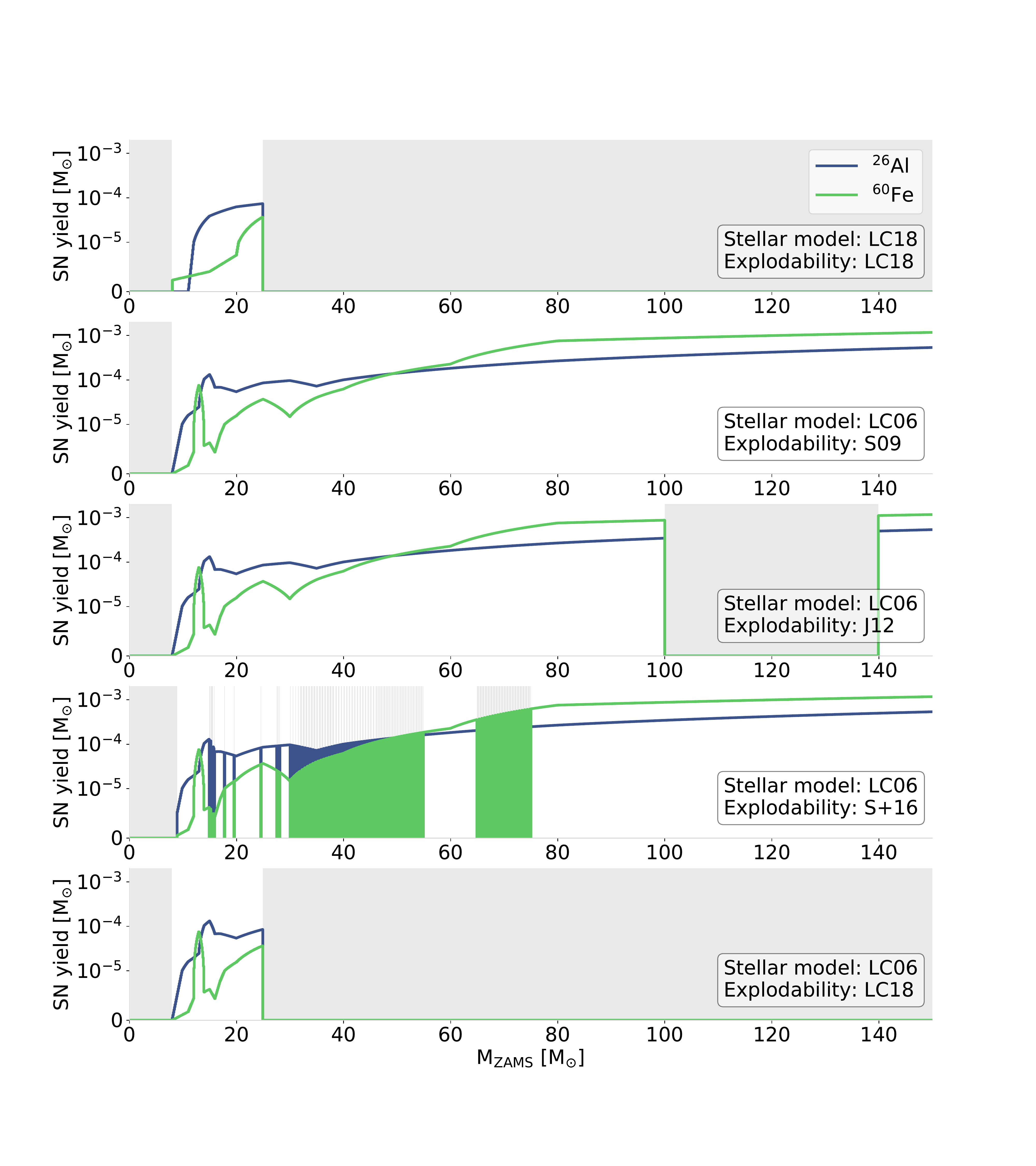}
	\caption{\Al (\emph{blue}) and \Fe (\emph{green}) yields from \acp{SN} by \citet{Limongi:2006gh} for different explodability models. Stars with an initial mass inside the grey shaded regions eject no material during the SN. Islands of explodability following each other closely appear as green regions.}
	\label{fig:explodability}
\end{figure}

At the end of the evolution of massive stars, a lot of processed material is ejected in \acp{SN}.
However, this only applies if the stellar collapse is actually followed by an explosion, which is parametrised as `explodability'.
Explosions can be prevented under certain circumstances if the star collapses directly into a black hole instead.
Depending on the complex pre-\ac{SN} evolution, this naturally has a strong impact on nucleosynthesis ejecta.
Different simulation approaches by \citet[][S09]{Smartt:2009cc}, \citet[][J12]{Janka:2012ex}, \citet[][S+16]{Sukhbold:2016aa}, or \citet[][LC18]{Limongi:2018gh} provide strikingly different explodabilities.
Fig.\,\ref{fig:explodability} shows effects on \Al and \Fe ejection, respectively, over the entire stellar mass range with nuclear yield calculations by \citet[][LC06]{Limongi:2006gh}.
While the \ac{SN} yields cease with suppression of the explosion, the wind ejecta remain unaffected by explodability.
Because \Fe is ejected only in \acp{SN} but \Al also in winds, the \Fe/\Al ratio is an important tracer of explodability effects on chemical enrichment.
The explodability can thus be chosen in the model as input parameter in order to test their astrophysical impact on large-scale effects of nucleosynthesis ejecta.

\subsubsection{Nucleosynthesis Yields}\label{sec:nucsys}
The most fundamental input to the modelling of nucleosynthesis ejecta is the total mass produced by each star over its lifetime.
This yield depends on many stellar factors, such as rotation, mixing, wind strength, metallicity, etc., as described above.
It also involves detailed nuclear physics, which is represented in the nuclear reaction networks of the stellar evolution models.

\begin{figure}[t]
	\centering
	\includegraphics[width=\hsize]{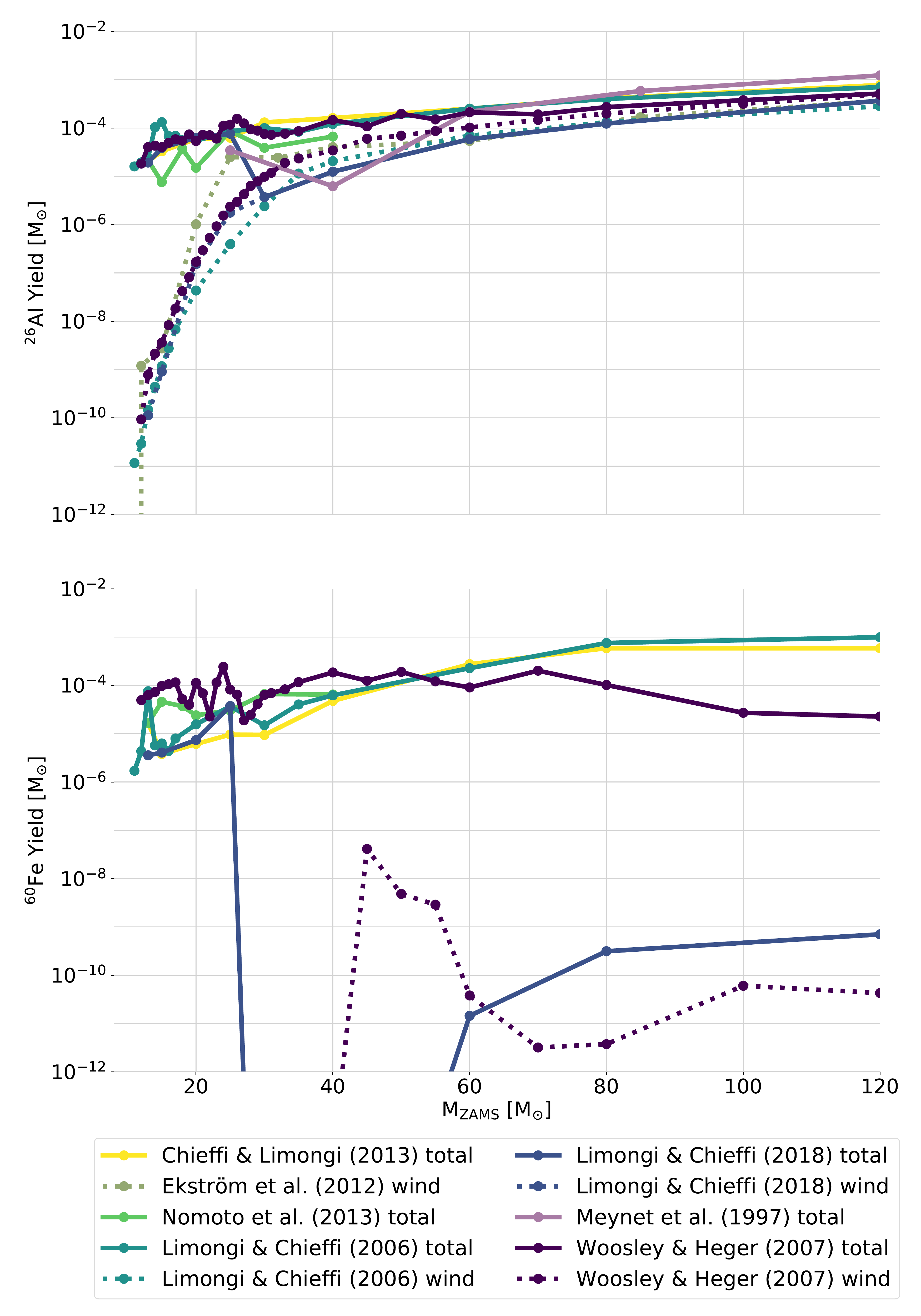}
	\caption{Nucleosynthesis yields (top \Al, bottom \Fe) for a selection of stellar evolution models, separated into the stellar wind and total contribution (wind + \ac{SN}).}
	\label{fig:yields}
\end{figure}

Detailed yield calculations for \Al and \Fe in particular have been performed e.g.\ by \citet{Meynet:1997al}, \citet{Limongi:2006gh}, \citet{Woosley:2007al}, \citet{Ekstrom:2012al}, \citet{Nomoto:2013al}, \citet{Chieffi:2013al}, or \citet{Limongi:2018gh}.
A comparison of the models is shown in Fig.\,\ref{fig:yields}.
While the \Al yield is generally dominated by \ac{SN} ejection in the lower-mass range, it is outweighed by wind ejection in the higher-mass regime.
The \Fe wind yields are negligible.
On average, a massive star ejects on the order of $10^{-4}$\,\Msol of \Al as well as \Fe.
The predictions show a spread of about one to two orders of magnitude between models in the stellar mass range of 10--120\,\Msol.
Stars of lower mass, such as \ac{AGB} stars with about 4\,\Msol are expected to eject much less \Al around $\sim 5 \times 10^{-5}$\,\Msol \citep{Bazan:1993al}.

For the overall contribution to nucleosynthesis feedback it is important to take the formation frequency of stars in a certain mass range into account.
A convolution with the \ac{IMF} shows that stars that form with a \ac{ZAMS} mass of $\leq30$\,\Msol contribute overall more \Al to the \ac{ISM} than more massive stars.
Due to the sensitivity of \Fe to explodability, the contribution to its overall amount by stars with $M_{\text{\ac{ZAMS}}} \geq 25$\,\Msol is negligible if \acp{SN} are not occurring. 
In our galactic nucleosynthesis model, we use the stellar evolution models by \citet[][LC06]{Limongi:2006gh} and \citet[][LC18]{Limongi:2018gh} as they include \Al and \Fe production and ejection in time-resolved evolutionary tracks over the entire lifetime of the stars, thus avoiding the need for extensive extrapolations.
We thus obtain population synthesis models of star groups that properly reflect the underlying physical feedback properties and their time variability.
Other yield models can also be included, if a large enough grid in the required model parameters is available, so that interpolations can rely on a densely- and regularly-spaced input.
We note again that individual yields of radioactive isotopes are quite uncertain as they depend on extrapolations from laboratory experiments towards different energy ranges.
For example, \citet{Jones2019_60Fe_crosssection} found that the cross section $\mathrm{^{59}Fe(n,\gamma)^{60}Fe}$ has a linear impact on the \Fe yields and suggest a smaller than previously accepted value to match \gray observations.

\subsection{Stellar binaries}\label{sec:binaries}
The impact of binary star evolution, especially in terms of nucleosynthesis, is a heavily debated field.
Roche-lobe overflows and tidal interactions can change the composition of binary stars as a whole and also enhance the ejection of material.
The effects of binary star evolution are highly complex because of the unknown influence of the many stellar evolution parameters.
First binary yield calculations for \Al have been provided by \citet{Brinkman:2019bi}.
We perform a quantitative check of binarity impacts on the stellar group scale by including these yields in \ac{PSYCO}.

\begin{figure}[t]
    \centering
    \includegraphics[width=\columnwidth]{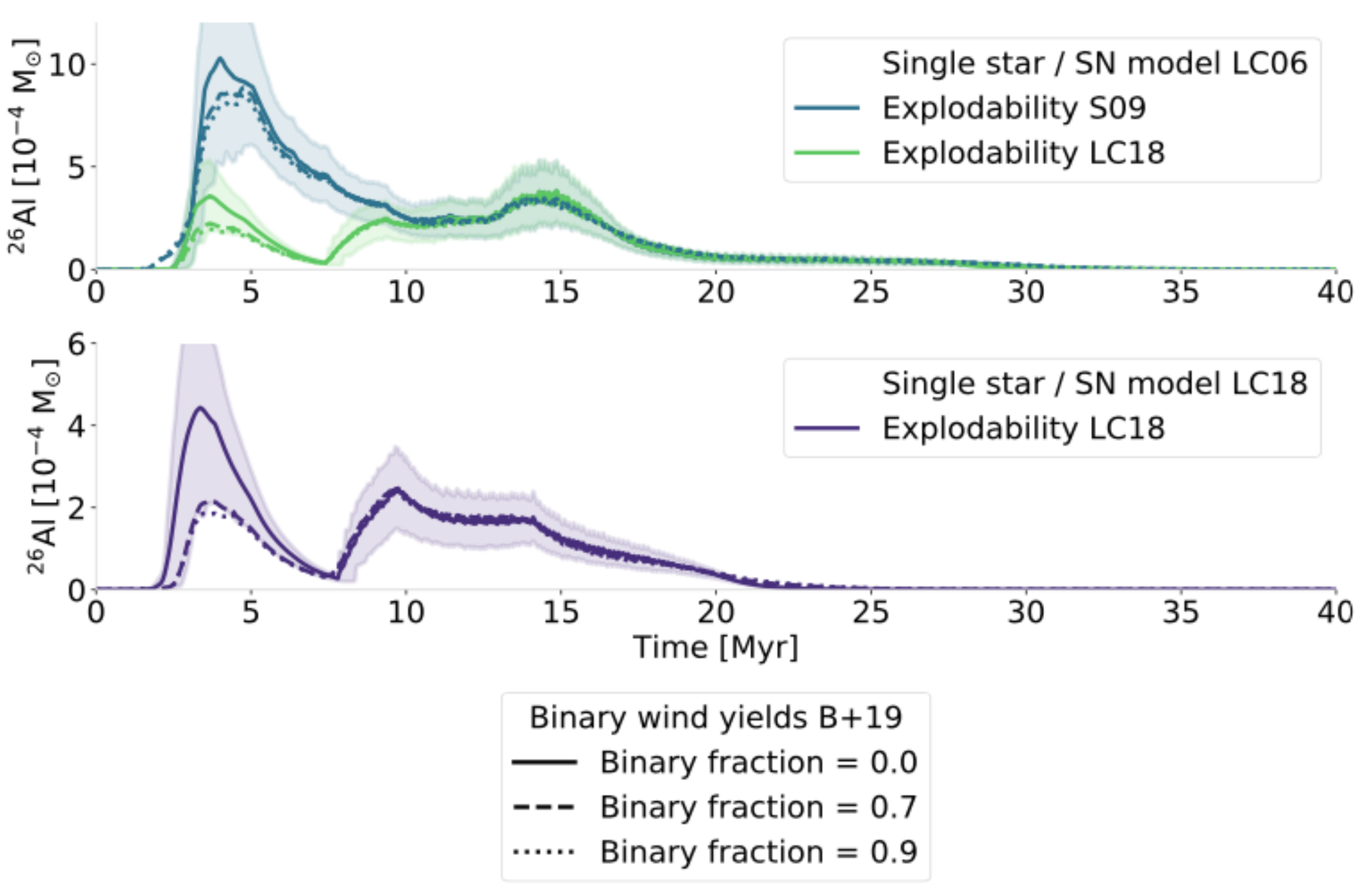}
    \caption{Population synthesis a canonical star group of $10^4$\,Msol including binary effects. Shown is the impact of \Al binary wind yields from \citet[][B+19]{Brinkman:2019bi}. Binary systems are included with orbital periods between 3 and 100 days and with an overall fraction of 70\,\% (dashed line) and 90\,\% (dotted line), which is contrasted with an association of only single stars (solid line). Binaries with longer periods, i.e. wider separations are considered to evolve as single stars. Lines indicate the average of 1000 Monte Carlo runs each.}
    \label{fig:binary_yields}
\end{figure}

For a canonical stellar group of $10^4$\,\Msol, we assume an overall binary fractions of 70\,\% \citep{Sana:2012bi,Renzo:2019bi,Belokurov:2020bi}, as well as extreme values of 0\,\% and 90\,\%.
If a star has a companion or not is sampled randomly according to the selected fraction as they emerge from the same \ac{IMF}.
\citet{Brinkman:2019bi} restricted the stellar evolution to that of the primary star so that we treat the companion as a single star.
In addition, \citet{Brinkman:2019bi} only considers wind yields so that we have to assume SN ejecta to follow other models.
We inter- and extrapolate the parameter grid from \citet{Brinkman:2019bi}, including orbital periods, masses, and orbit separations to a similar grid as described above.
The resulting population synthesis for a single stellar group is shown in Fig.\,\ref{fig:binary_yields}.
The two extreme explodability assumptions S09 and LC18 are shown for comparison.
Independent of the SN model choice, the effects from binary wind yields appear rather marginal.
In particular for low-mass stars, being the dominant \Al producers after 10\,Myr, considering binaries has no impact.
The reduction of wind yields in binary systems with large separation and primary stars of 25--30\,\Msol leads to less \Al ejection after 15\,Myr.
The increased binary wind yields for stars with $\lesssim 20$\,Msol results in a slightly enhanced ejection after 17\,Myr.
In addition, at early stages, the ejection from very massive stars tends to be reduced due to binary interactions.
It is important to note here that these extra- and interpolations come with large uncertainties so that the binary yield considerations here should not be overinterpreted.
Given the wind yield models by \citet{Brinkman:2019bi}, the variations for a canonical star group all lie within the 68th percentile of a single star population synthesis results.

\section{Population Synthesis}\label{sec:popsyn}
The cumulative outputs of the \ac{PSYCO} model is built up step by step using the input parameters outlined in Sec.~\ref{sec:model_params} as depicted on the left in Fig.~\ref{fig:psyco_structure} from the yields of single stars to properties of massive star groups to the entire Galaxy.
The underlying method is a population synthesis approach, which relates the integrated signal of a composite system with the evolutionary properties of its constituents \citep{Tinsley:1968ps, Tinsley:1972ps, Cervino:2013ps}.

\subsection{Star Group}\label{sec:star_group}
Individual sources of interstellar radioactivity remain unresolved by current \gray instruments.
However, integrated cumulative signals from stellar associations can be observed \citep[e.g.,][]{Oberlack1995_Orion26Al,Knoedlseder1996_26Al,Kretschmer:2000wg, Knodlseder:2000jw, Martin:2008yt, Martin:2010ko, Diehl:2010bc, Voss:2010et, Voss:2012aa}.
This describes star groups as the fundamental scale on which the Galactic model of nucleosynthesis ejecta is based upon.

\begin{figure*}[t]
	\subfloat[Variations with different IMFs.]{
		\begin{minipage}[b][0.7\width]{0.5\textwidth}
			\centering
			\includegraphics[width=1\textwidth, trim=2.5cm 10cm 2.5cm 4.5cm, clip]{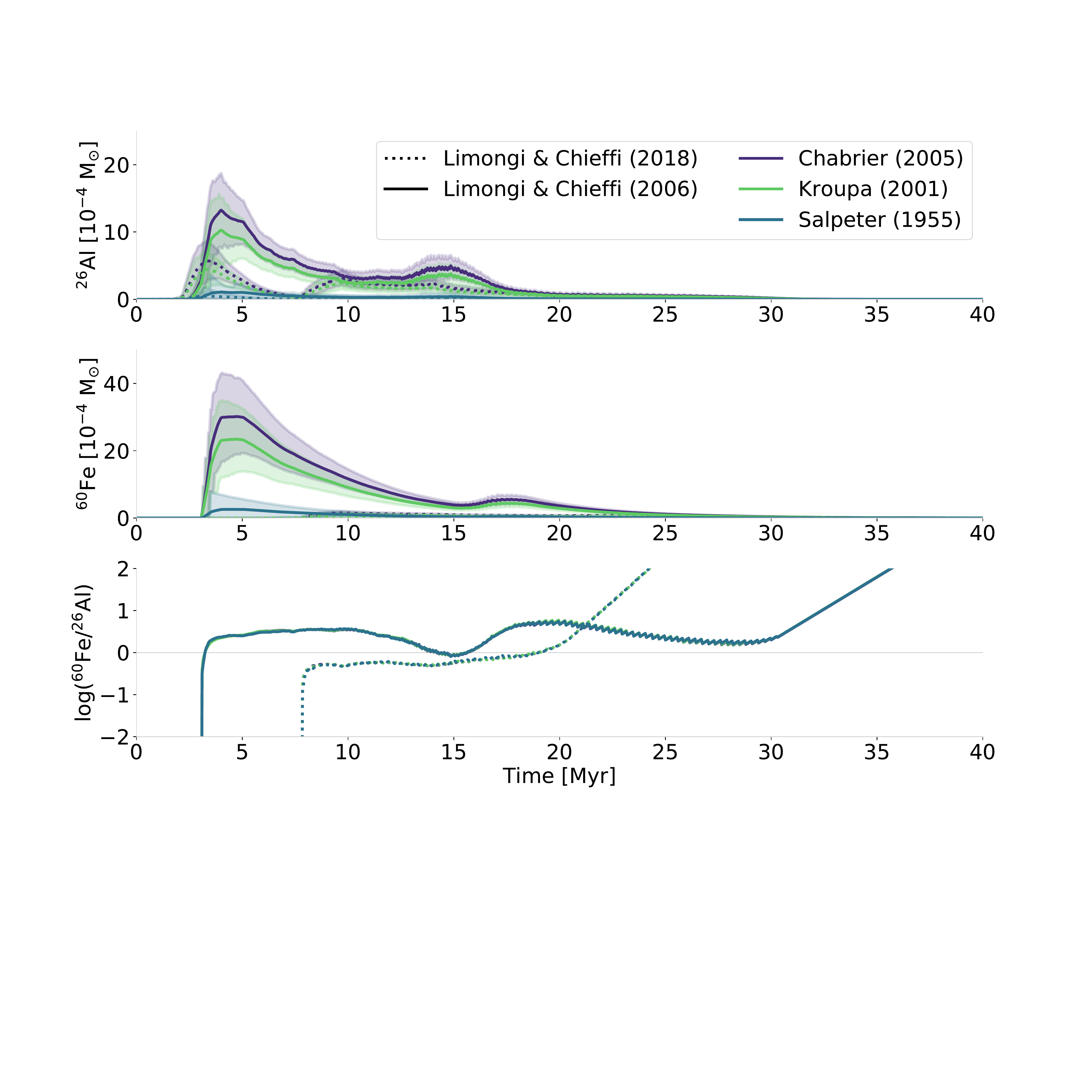}
	\end{minipage}}
	\subfloat[Variations with different metallicities.]{
		\begin{minipage}[b][0.7\width]{
				0.5\textwidth}
			\centering
			\includegraphics[width=1\textwidth, trim=2.5cm 11cm 2.5cm 4.5cm, clip]{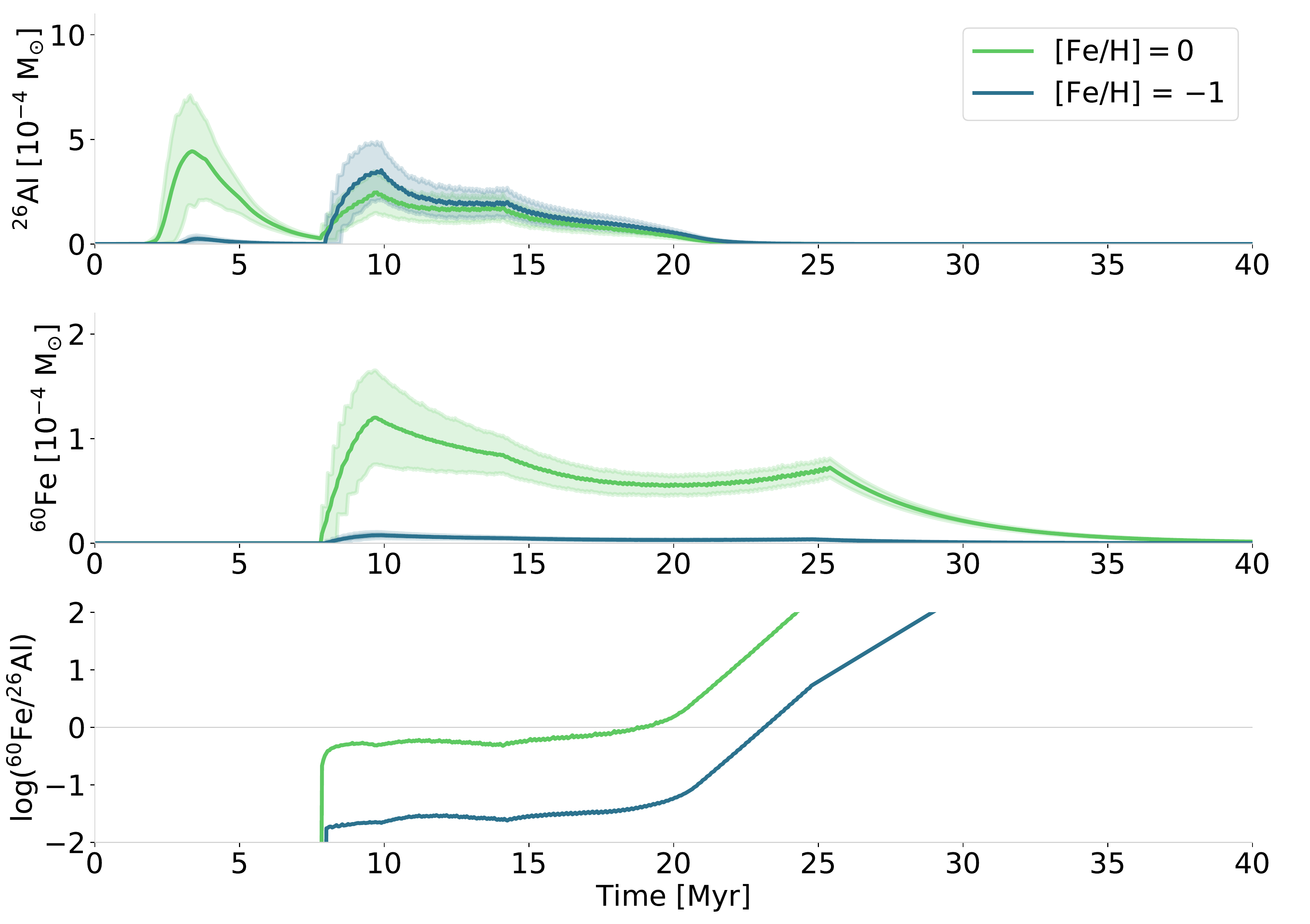}
	\end{minipage}}\\
	\subfloat[Variations with different explodabilities.]{
		\begin{minipage}[b][0.7\width]{
				0.5\textwidth}
			\centering
			\includegraphics[width=1\textwidth, trim=2.5cm 11cm 2.5cm 4.5cm, clip]{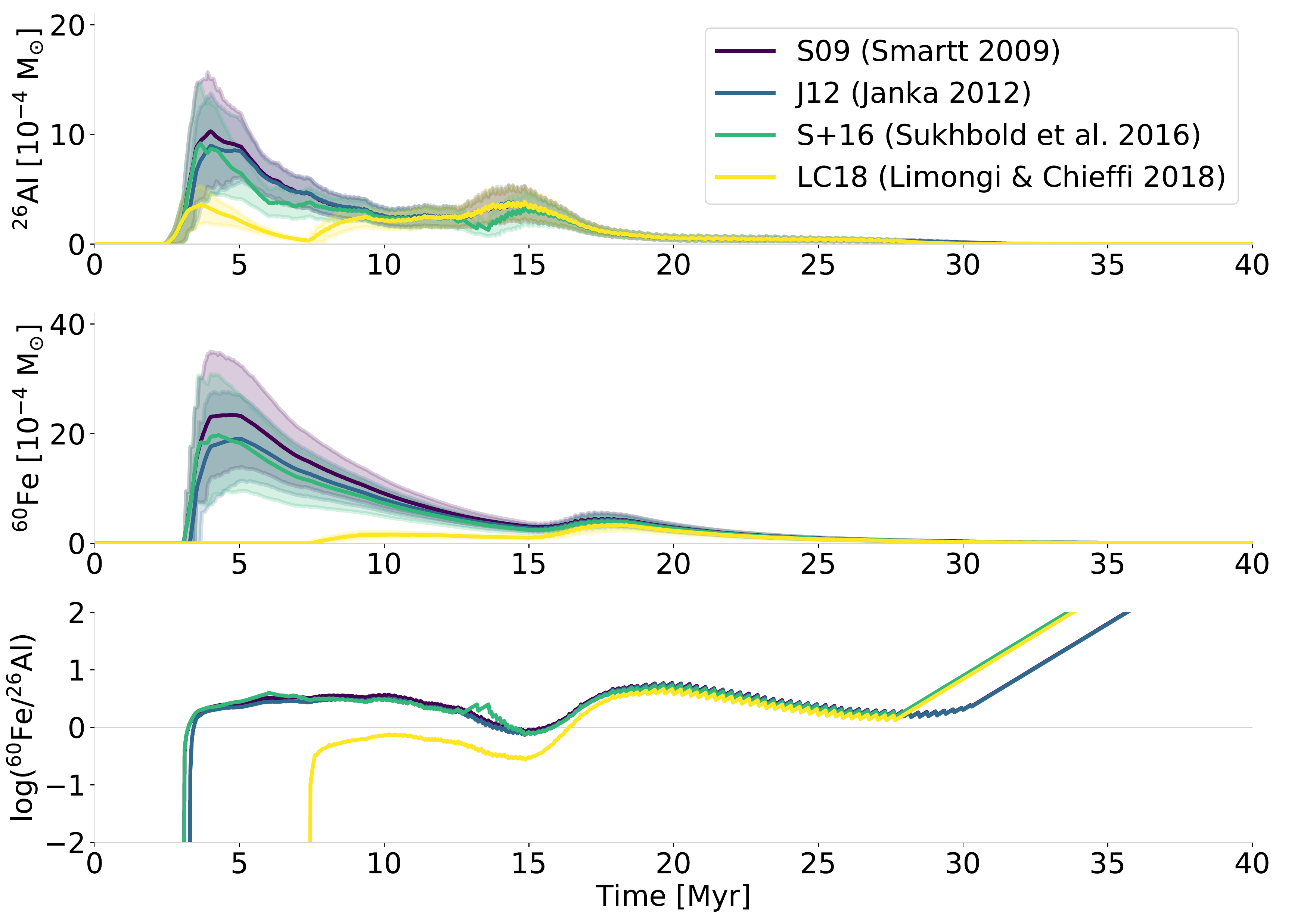}
	\end{minipage}}
	\subfloat[Difference when including rotation.]{
		\begin{minipage}[b][0.7\width]{
				0.5\textwidth}
			\centering
			\includegraphics[width=1\textwidth, trim=2.5cm 11cm 2.5cm 4.5cm, clip]{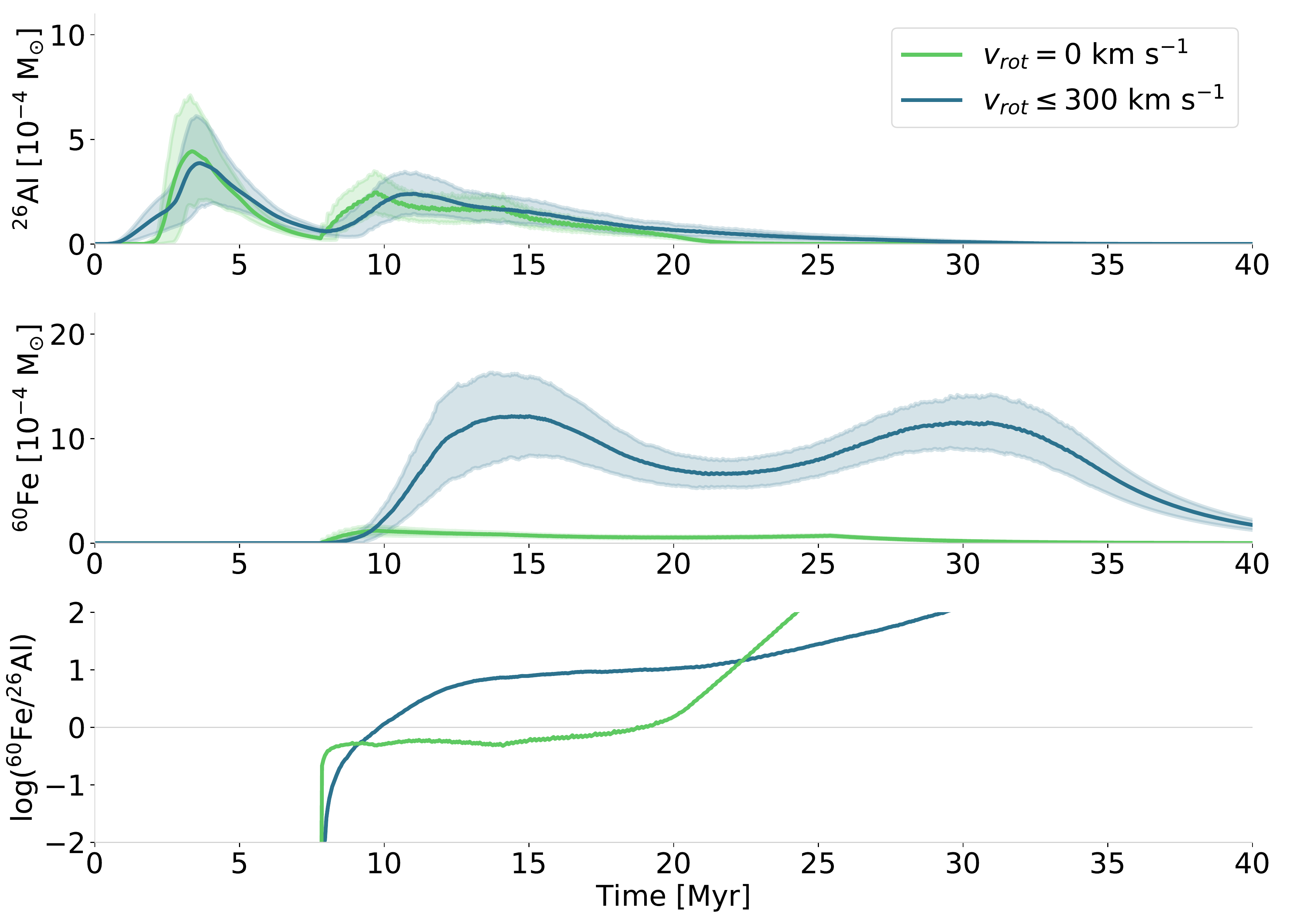}
	\end{minipage}}
	\caption{Population synthesis models of nucleosynthesis ejecta in a $10^{4}$\,\Msol star group based on yield models by \citet[][LC06]{Limongi:2006gh} or \citet[][LC18]{Limongi:2018gh} as indicated in subcaptions. Different panels illustrate the impact of the physical parameters (a) \ac{IMF}, (b) metallicity, (c) explodability, and (d) stellar rotation. Each panel shows the time evolution of the ejection of \Al (\emph{upper}), \Fe (\emph{middle}), and their average mass ratio (\emph{bottom}). In each case, coeval star formation is assumed and the shaded regions indicate the 68th percentiles derived from $10^{3}$ Monte Carlo runs.}
	\label{fig:popsyn_stellar_group}
\end{figure*}

In massive star population synthesis, time profiles $\psi(M_*, t)$  of stellar properties (e.g., ejecta masses, UV brightness) are integrated over the entire mass range of single stars, weighted with the IMF $\xi(M_*)$,
\begin{equation} \label{eq:massive_star_population_synthesis}
\Psi(t) = A\int_{M_{*,\,\text{min}}}^{M_{*,\,\text{max}}}\psi(M_*, t)\ \xi(M_*)\ dM_*,
\end{equation}
with the normalisation $A$ according to the total mass of the population (group or cluster).
It has been shown that continuous integration of the \ac{IMF} within population synthesis calculations can result in a considerable bias for cluster properties such as its luminosity \citep{Piskunov:2011ps}.
This is related to the discrete nature of the \ac{IMF}, resulting in a counting experiment with seen and unseen objects of a larger population, which is difficult to properly treat without knowing the selection effects.
In order to take stochastic effects in smaller populations into account, we therefore use a discrete population synthesis by \ac{MC} sampling \citep{Cervino:2006ps}.
Publicly accessible population synthesis codes have been developed and applied successfully to astrophysical questions \citep[e.g.,][]{Popescu:2009ps,Silva:2012aa}.
A generic population synthesis code including selection effects, biases, and overarching distributions has recently been developed by \citet{Burgess2021_popsynth}.
Focussing on the specific case of radionuclei and interstellar \gray emission, we base our population synthesis approach on the work by \citet{Voss:2009iq}, who also included kinetic energy and UV luminosity evolution of superbubbles.

In a first step, individual initial mass values are sampled according to the \ac{IMF}.
In order to assure a discretisation and to reproduce the shape of the \ac{IMF} as close as possible, we apply the optimal sampling technique, which was developed by \citet{Kroupa:2013bw} and revised as well as laid out in detail by \citet{Schulz:2015ec}.
\citet{Krause:2015aa} have shown that the details of the sampling method do not influence \nuc{Al}{26} abundances and superbubble properties significantly.
Details of the optimal sampling method is given in appendix\,\ref{sec:appendix_optimal_sampling}.
It is based on the total mass $M_{\rm EC}$ of the cluster to be conserved,
\begin{equation} \label{eq:optimal_sampling_total_mass}
M_{\text{EC}} = \int_{M_{\text{min}}}^{M_{\text{max}}} M_*\ \xi(M_*)\ dM_*,
\end{equation}
during the formation of stars with mass $M_{\text{min}} \leq M_* \leq M_{\text{max}}$ according to an \ac{IMF} $\xi(M_*)$.

In the next step, each single star is assigned a stellar rotation velocity according to its spectral class (Sect.\,\ref{subsubsec:stellar_rotation}).
In addition to the total mass, each star group is assigned a position drawn from the Galactic density distribution (Sect.\,\ref{subsubsec:spatial_characteristics}) and accordingly a metallicity (Tab.\,\ref{tab:metallicity_gradient}).

By adding the respective stellar isochrones, we thus obtain cumulative properties of star groups based on the described parameters and assuming coeval star formation for any given event/group.
The original evolutionary tracks by \citet{Limongi:2006gh} and \citet{Limongi:2018gh} cover only a few \ac{ZAMS} masses and irregular time steps.
For the population synthesis, a uniform and closely-meshed (fine) grid of stellar masses in $0.1$\,\Msol steps and evolution times in $0.01$\,Myr is created by interpolations.
We use linear extrapolation to include stellar masses above $120$\,\Msol and below $13$\,\Msol.
Fig.\,\ref{fig:popsyn_stellar_group} shows the effect of the main physical input parameters on \Al and \Fe ejection with a population synthesis of a $10^{4}$\,\Msol cluster for different assumptions of the \ac{IMF}, metallicity, explodability, and stellar rotation.
They are mainly based on models by \citet{Limongi:2018gh}, as they cover this whole range of physical parameters.
Models by \citet{Limongi:2006gh} are chosen to show explodability effects, as they do not include intrinsic assumptions about this parameter.
For better cross-comparison, stellar masses have been obtained by random sampling in this case to determine stochastic uncertainty regions.

The choice of the \ac{IMF} (Fig.\,\ref{fig:popsyn_stellar_group}, top left) affects the amplitude of nucleosynthesis feedback for both \Al and \Fe.
While K01 and C05 yield similar results within the statistical uncertainties, S55 shows a strong reduction of nucleosynthesis ejecta.
This is readily understood because S55 continues unbroken towards low-mass stars which is unphysical.
This distributes a great amount of mass into a large number of low-mass stars -- i.e. those which do not produce major amounts of \Al or \Fe.

The production of \Al and \Fe generally relies strongly on the initial presence of seed nuclei, mainly \element[][25]{Mg} and \element[][59]{Fe}, respectively.
Thus, an overall metallicity reduction in the original star forming gas ultimately also goes along with a decrease of ejection yields of \Al and \Fe (Fig.\,\ref{fig:popsyn_stellar_group}, top right) \citep{Timmes:1995al,Limongi:2006gh}.
This effect is strong for \Fe because \element[][56]{Fe} is produced only on a very short time scale in late evolutionary stages.
Thus, only marginal amounts of processed \element[][56]{Fe} can reach the He- or C-burning shells where \Fe production can occur \citep{Tur:2010fe,Uberseder:2014fe}.
Reduced metallicity also decreases the opacity of stellar material \citep{Limongi:2018gh}.
As a consequence, convective zones shrink and stellar winds decrease with lower radiative pressure.
Both effects reduce \Al yields because it resides in hot regions where it is destroyed and the wind component ceases.

The impact of explodability is shown to be strongest for \Fe because this isotope is ejected only in \acp{SN} (Fig.\,\ref{fig:popsyn_stellar_group}, bottom left).
The more extensive the inhibition of explosions, the stronger the reduction of \Fe yields, especially at early times (higher initial masses).
This effect is comparably weak for \Al and accounts for only a factor of 2 less ejection from the most massive stars because the wind component remains unaffected.

Due to the increased centrifugal forces in fast rotating stars (Fig.\,\ref{fig:popsyn_stellar_group}, bottom right), the core pressure is reduced and its overall lifetime extended \citep{Limongi:2018gh}.
Thus, nucleosynthesis feedback is stretched in time when stellar rotation is included.
For \Al, this effect is mainly recognisable as a slightly earlier onset of winds.
The changes in \Fe are much stronger.
The enhancement of convection zones increases the neutron-rich C- and He-burning shells significantly and enhanced material from even deeper layers can be mixed into these regions. This leads to a boost in \Fe production by a factor of up to 25.
Additionally, the duration of \Fe ejection in a star group is prolonged by a factor of about 2.

Fig.\,\ref{fig:popsyn_stellar_group} shows that the mass ratio \Fe/\Al is particularly sensitive to changes in metallicity, explodability and rotational velocity.
A one order of magnitude reduction in metallicity leads to a decrease of the same magnitude in this mass ratio.
The most significant impacts on the temporal behaviour have rotation effects.
Due to the prolongation of stellar evolution and the drastic increase in \Fe ejection, \Fe/\Al is dominated by \Fe after only $\sim10$\,Myr.
If stellar rotation is not taken into account \Al dominance lasts for $\sim18$\,Myr.
Changes in explodability show also a shift in this time profile.
If explosions of massive stars with $M_* > 25$\,\Msol are excluded, \Fe domination is delayed by $\sim13$\,Myr to about 16\,Myr after cluster formation.
This underlines that the \Fe/\Al ratio is an important observational parameter that provides crucial information about detailed stellar physics.

Due to the reproducibility and uniqueness of optimal sampling, time profiles of star groups can be calculated in advance.
We take advantage of this fact and create a database that covers a broad parameter space of combinations of cluster masses, explodability models, yield models and IMF shapes.
This procedure drastically reduces the computing time of the overall galaxy model (Sect.\,\ref{sec:galaxy}).

\subsection{Galaxy}\label{sec:galaxy}
We extend the population synthesis to the galactic level by calculating a total galactic mass that is processed into stars with a constant star formation rate over 50\,Myr as described in Sect.\,\ref{sec:stellar_mass}.
Because the \ac{ECMF} behaves similarly to the \ac{IMF}, the optimal sampling approach is also used here.
In addition to the variables of mass and time, the spatial dimension is added at this level.
Star groups form at different positions and times in the Galaxy and their spatial extents evolve individually over time (Eq.~\ref{eq:bubble_expansion}).
In order to transfer this 3D information into an all-sky map that can be compared to actual \gray measurements, we perform a line-of-sight integration (for details, see Appendix\,\ref{sec:appendix_los_integration}).

\begin{figure}[t]
	\centering
	\includegraphics[width=\hsize]{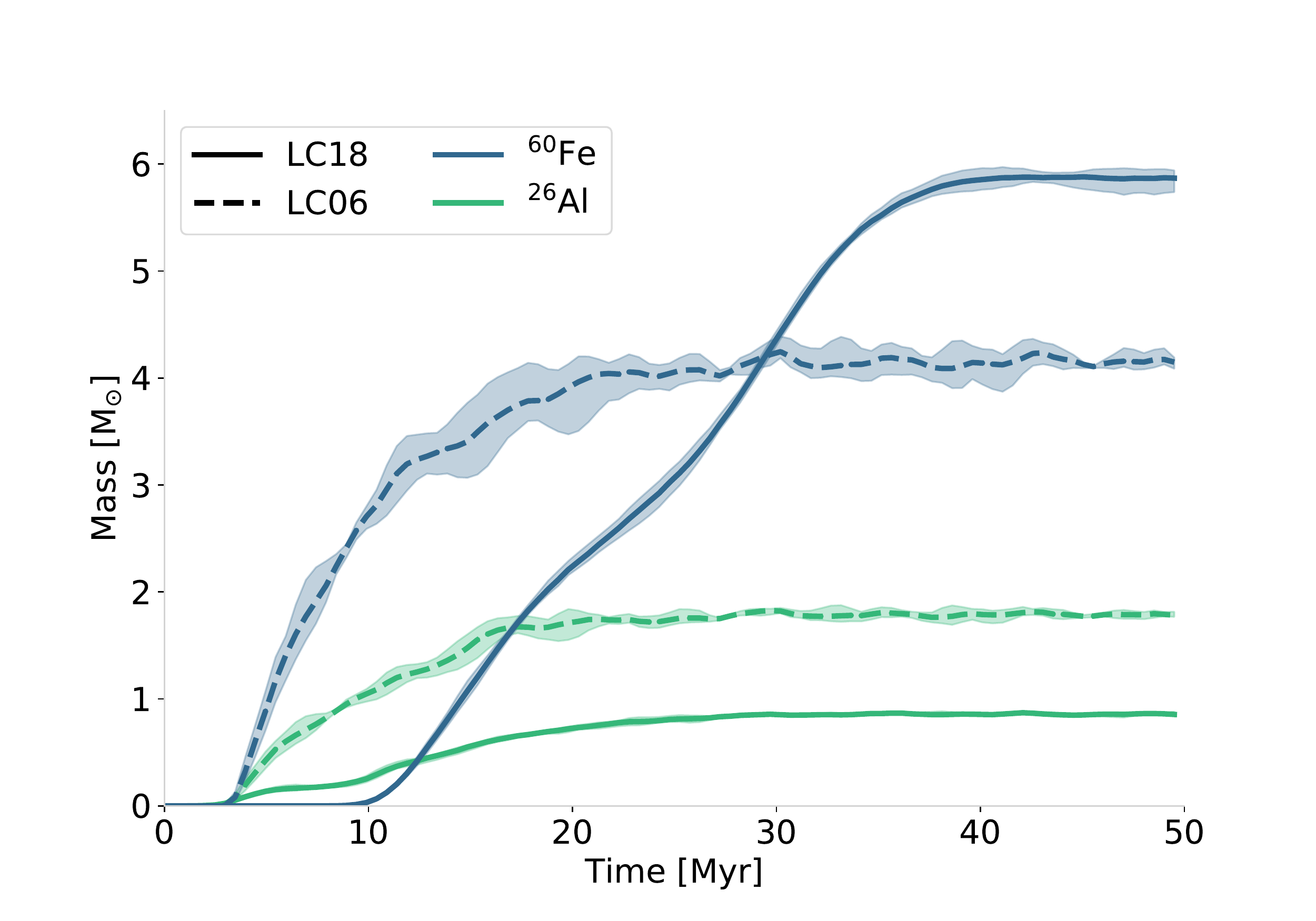}
	\caption{Steady-state settling time of the total \Fe (\emph{blue}) and \Al (\emph{green}) in \ac{PSYCO} galaxy models. Shaded regions denote the 68th percentile of 100 \ac{MC} model runs. All models are based on evolutionary tracks LC06 (dashed lines) or LC18 (solid lines) and explodability models S09 and LC18, respectively, for $\text{SFR} = 4\,\mathrm{M_{\odot}\,yr^{-1}}$ and the K01 \ac{IMF}.}
	\label{fig:time_sequence}
\end{figure}

The ejecta of an isotope $n$ with atomic mass $m_{n,\text{u}}$ distributed homogeneously in a bubble with radius $R_{\text{SB}}(t)$ at time $t$ emit \grays at energy $E_{\gamma}$ from a nuclear transition with probability $p_{E_{\gamma}}$ with a luminosity
\begin{equation}\label{eq:unit_luminosity}
L_{n, \odot} = \frac{L_n}{M_{\odot}} =  \frac{ p_{E_{\gamma}} }{ M_{m,\text{u}}  \tau_n},
\end{equation}
which is normalised to a unit Solar mass of isotope $n$ and is expressed in units of ph\,s$^{-1}$\,\Msol$^{-1}$.
For example in the case of \Al, the 1.8\,MeV luminosity per solar mass is $1.4 \times 10^{42}\,\mathrm{ph\,s^{-1}\,M_{\odot}^{-1}} = 4.1 \times 10^{36}\,\mathrm{erg\,s^{-1}\,M_{\odot}^{-1}}$.
The amount of isotope $n$ present in the superbubble is predetermined for each point in time by the massive star population synthesis.
This determines an isotope density $\rho_n$ which is constant inside and zero outside the bubble.
\citet{Burrows1993_OrionEridanus} and \citet{Diehl:2004ke} suggest that ejecta remain `inside' the bubble.
There could be some mixing of \Al with the HI walls.
Hydrodynamic simulations find a varying degree of concentration of nucleosynthesis ejecta in the supershells \citep{Breitschwerdt:2016ds,Krause:2018aa}.
In any case, the superbubble crossing time of about 1\,Myr again would make the ejecta appear homogenised.
Given the angular resolution of current \gray instruments of a few degrees we therefore find that a homogeneous density inside the bubbles is a good first-order approximation (see Sect.\,\ref{sec:caveats} for a discussion).

By line-of-sight integration of these homogeneously-filled spheres, a spatial \gray emission model for each superbubble is created and added onto the current model map.
Their cumulative effect finally gives a complete galactic picture.
This formulation is easily adaptable to arbitrary isotopes by scaling $L_{n, \odot}$ and lifetimes $\tau_n$.
In the case of short-lived isotopes such as $^{44}$Ti with a lifetime of 89\,yr, for example, the spatial modelling reduces to point sources for SPI because the ejecta do not travel far from their production site before decaying.

\begin{figure*}[t]
    \centering
    \includegraphics[width=\textwidth]{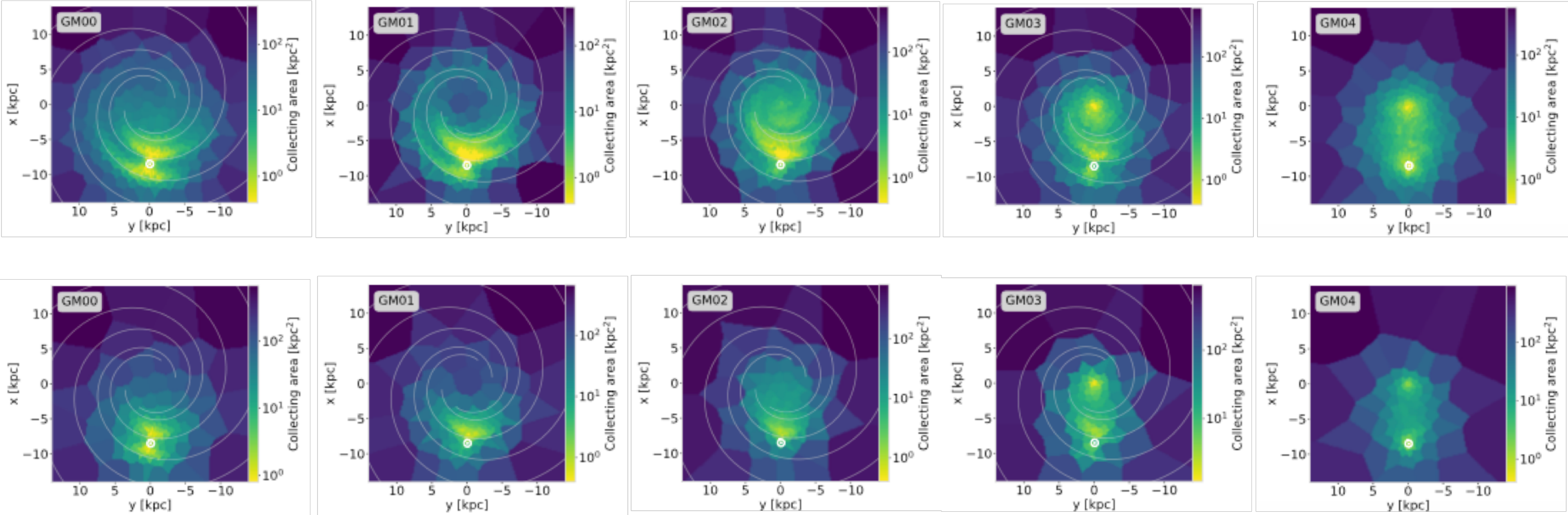}
    \caption{Radial distribution of modelled flux contributions for a theoretical observer (white circle) from \Al (\textit{top}) and \Fe (\textit{bottom}) decay. Each column represents average results of 500 model instantiations based on the density profiles GM00--GM04 (grey boxes). The models shown are stellar evolution models LC06, explodability S09, IMF K01, and $\mrm{SFR} = \Msol\,\mrm{yr^{-1}}$. The latter corresponds to a total mass of $1.8 \pm 0.2$\,\Msol and $4.2 \pm 0.2$\,\Msol of \Al and \Fe, respectively. Adaptive spatial binning \citep{Cappellari:2003vb}is used to obtain Voronoi tessellations as spatial bins, each of which contribute a flux of $10^{-6}$\,\flux for the observer. The colour scale refers to the collecting area covered by each such pixel.}
    \label{fig:radial_profiles}
\end{figure*}

\section{Simulation results}\label{sec:performance}
\subsection{Evaluation of PSYCO Models}\label{sec:psyco_model_comparisons}
We have evaluated a grid of models, varying our input parameters with $\mrm{SFR} \in \{2,4,8\}$\,\Msol\,$\mrm{yr^{-1}}$, scale height $z_0 \in \{0.1, 0.2, 0.3, 0.5, 0.7\}$\,kpc, density profiles GM00--GM04, the two stellar evolution models LC06 and LC18, and the explodabilities S09, and S+16 (and LC18 to match the LC18 stellar evolution model).
We chose to use only the \ac{IMF} K01.
For each parameter value combination, 100 \ac{MC} runs are performed to estimate stochastic variations, which in total amounts to 30000 simulated \ac{PSYCO} maps.

From this number of simulations, we can explore links (correlations) between the parameters and assign some uncertainty to those.
Naturally, the \ac{SFR} and explodability have an impact on the total amount of \Al and \Fe present in Galaxy:
For LC18 (stellar evolution model and explodability), the total galactic \Al mass follows roughly a linear trend $M_{26}/\Msol \approx 0.25 \times \mrm{SFR/(\Msol\,yr^{-1})}$; for other explodabilities, the \ac{SFR} impact is larger $M_{26}/\Msol \approx 0.31$--$0.52 \times \mrm{SFR/(\Msol\,yr^{-1})}$.
For \Fe, the effects of explodability are reversed since \Fe is only ejected in SNe.
We find $M_{60}/\Msol \approx 1.72 \times \mrm{SFR/(\Msol\,yr^{-1})}$ for LC18, and $M_{60}/\Msol \approx 0.28$--$1.27 \times \mrm{SFR/(\Msol\,yr^{-1})}$ for the other explodability models.
The resulting mass ratio \Fe/\Al has therefore almost no \ac{SFR}-dependence, and we find \Fe/\Al of 0.9 for LC18, up to 7.1 for LC06.
We note that there are crucial differences in the flux, mass, and isotopic \Fe/\Al ratio:
Given that the \gray flux $F_n$ of an radioactive isotope $n$ is proportional to $M_n p_{\gamma,n} m_n^{-1} \tau_n^{-1}$ (see Eq.\,(\ref{eq:unit_luminosity})), the flux ratio of \Fe to \Al in the Galaxy as a whole is
\begin{equation}\label{eq:flux_ratio}
        \frac{F_{60}}{F_{26}} = \frac{p_{60}}{p_{26}} \cdot \frac{\tau_{26}}{\tau_{60}} \cdot \frac{m_{26}}{m_{60}} \cdot \frac{M_{60}}{M_{26}} = 1.00 \cdot 0.27 \cdot 0.43 \cdot \frac{M_{60}}{M_{26}} = 0.12 \frac{M_{60}}{M_{26}}\mrm{.}
\end{equation}
The conversions between flux ratio and isotopic ratio, mass ratio $M_{60}/M_{26}$, isotopic ratio $N_{60}/N_{26} = M_{60}/M_{26} \cdot \frac{m_{26}}{m_{60}}$, and production rate $\dot{M}_{60}/\dot{M}_{26} = M_{60}/M_{26} \cdot \frac{\tau_{26}}{\tau_{60}}$, are given in Tab.\,\ref{tab:conversions}.
\begin{table}[h]
	\caption{Conversions between \Fe/\Al flux, mass, isotopic, and production ratios:}             % title of Table
	\label{tab:conversions}      % is used to refer this table in the text
	\centering                          % used for centering table
	\begin{tabular}{c | c c c c}        % centered columns (4 columns)
		 & $F_{60}/F_{26}$ & $M_{60}/M_{26}$ & $N_{60}/N_{26}$ & $\dot{M}_{60}/\dot{M}_{26}$ \\
		\hline
		$F_{60}/F_{26}$             & 1.00 & 0.12 & 0.27 & 0.43 \\
		$M_{60}/M_{26}$             & 8.43 & 1.00 & 2.31 & 3.65 \\
		$N_{60}/N_{26}$             & 3.65 & 0.43 & 1.00 & 1.58 \\
		$\dot{M}_{60}/\dot{M}_{26}$ & 2.31 & 0.27 & 0.62 & 1.00 
	\end{tabular}
\end{table}

The SN rates (SNRs) from these model configurations are directly proportional to the \ac{SFR}, as expected, and follow the trend $\mrm{SNR}/\mrm{century^{-1}} \approx 0.37$--$0.56 \times \mrm{SFR/(\Msol\,yr^{-1})}$, with the explodability LC18 giving the lowest SNR and S09 the highest.
The values above are independent of the chosen density profiles.
By contrast, the 1.809\,MeV (\Al) and 1.173 and 1.332\,MeV (\Fe) fluxes are largely dependent on the chosen spiral-arm prominence.
We show trends of these values as derived from PSYCO simulations in Appendix\,\ref{sec:appendix_PSYCO_trends}.

\subsection{Overall appearance}\label{sec:appearance}
Fig.\,\ref{fig:time_sequence} shows the convergence of \Fe and \Al masses in the model within $T_{\text{tot}} = 50$\,Myr.
This is an artificial diagnostic of radioactive masses to reach a constant value in a steady state.
The stellar evolution models LC06 and LC18 both lead to an equilibrium between production and decay within this time span owing to the constant star formation.
A change in \ac{SFR} alters the overall amplitude, leaving the general convergence behaviour unaffected.
This means that after a modelling time of 50\,Myr, the distribution of the isotopes \Al and \Fe is determined only by the assumed distribution of the star groups in space and time, and no longer by the initial conditions.
Due to the specific scientific focus on nucleosynthesis ejecta, we therefore choose the snapshot at $50\,\mrm{Myr}$ to evaluate all our models (see also Sect.\,\ref{sec:psyco_model_comparisons}).

About 50\,\% of the total \Al $\gamma$-ray flux is received from within 6\,kpc in GM03, i.e. the most shallow profile.
For the spiral-arm-dominated profiles, GM00 and GM01, half the flux is already contained within 2.8\,kpc.
It is important to note here that most of the flux received excludes the Galactic centre with a distance of 8.5\,kpc.
Until the distance of the Local Arm tangent at about 2\,kpc, on average 30\,\% of the flux is enclosed (see also Fig.\,\ref{fig:radial_profiles}).
In addition, it is interesting that in 0.3\,\% of all cases (i.e. 90 out of 30000 models, see Sect.\,\ref{sec:psyco_model_comparisons}), about 90\,\% of the total flux comes from a region of only 6\,kpc around the observer.
This means that the local components outweigh the overall Galactic emission by far in these cases.
More centrally dominated morphologies (especially GM04) show a flatter slope than spirally dominated ones (GM00--GM02).
Our models show similar flux profiles compared to the simulation by \citet{Rodgers-Lee:2019al}, for example.
With respect to this hydro-simulation, the best agreement is found with the centrally weighted spiral morphology GM03.
Particular observer positions in the hydrodynamic simulation can show a strong contribution from the Local Arm.
Such a behaviour is reflected by spiral arm dominated morphologies like GM01 in our model.
Based on these general agreements between galactic-wide population synthesis and hydrodynamics simulations, it is suggested that some important properties of the Galaxy can be transferred to the simpler-structured population synthesis model to test stochastic effects and a variety of parameters.

\begin{figure*}[p]
    \centering
    \includegraphics[width=0.85\textwidth,trim=0.1in 0.0in 0.0in 0.0,clip]{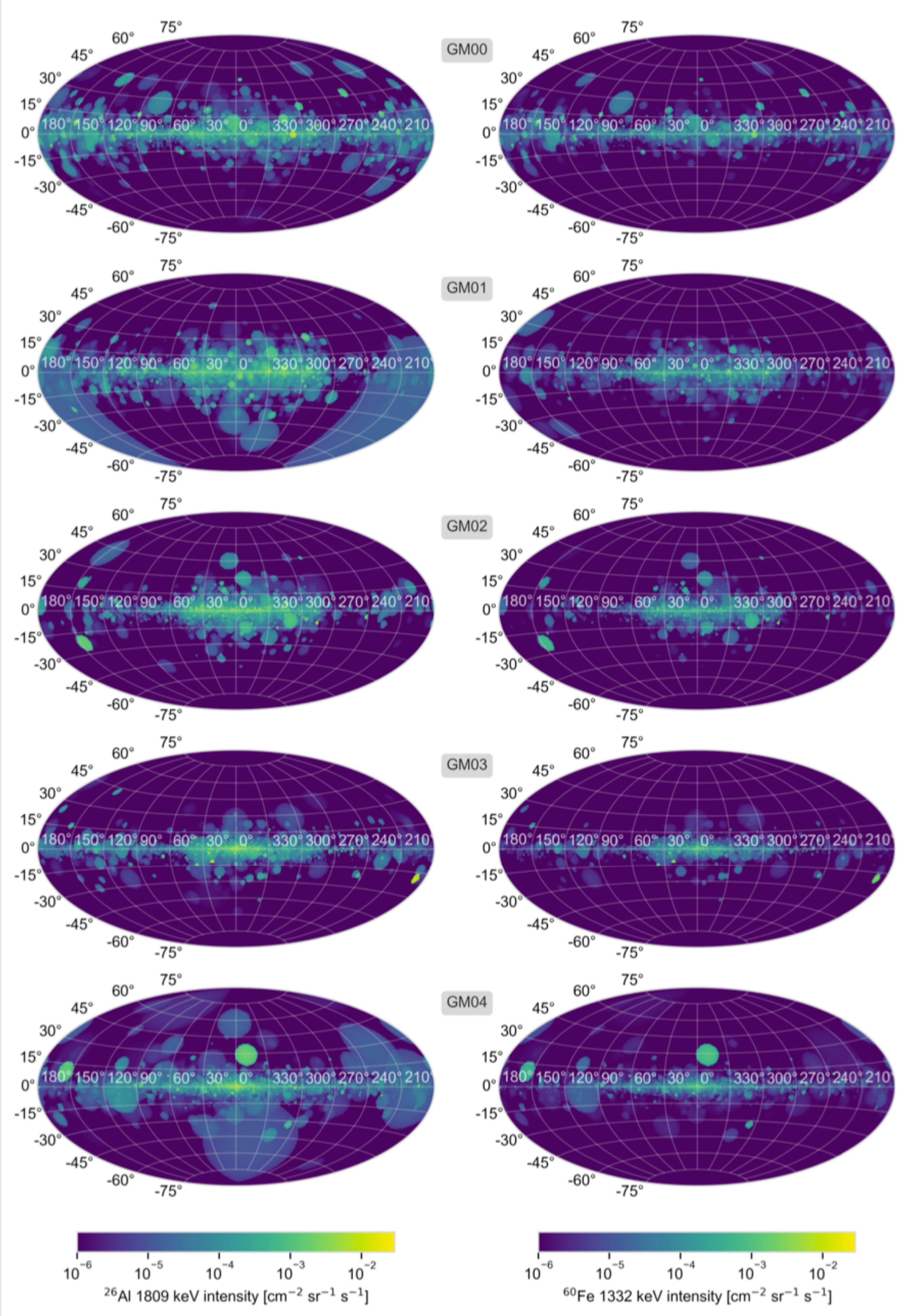}
    \caption{Simulated full-sky \gray maps of the 1.8\,MeV line from \Al decay (\textit{left}) and the 1.3\,MeV line from \Fe decay (\textit{right}) constructed with PSYCO. Each row represents an individual MC run based on a different density profile (gray boxes) with a scale height of 300\,pc. Nucleosynthesis yields are based on LC06 with the S09 explodability and the K01 IMF.}
    \label{fig:example_images}
\end{figure*}

\begin{figure*}[t]
    \centering
    \includegraphics[width=1.0\textwidth,trim=0.0cm 4.0cm 0.0cm 0.0cm,clip=True]{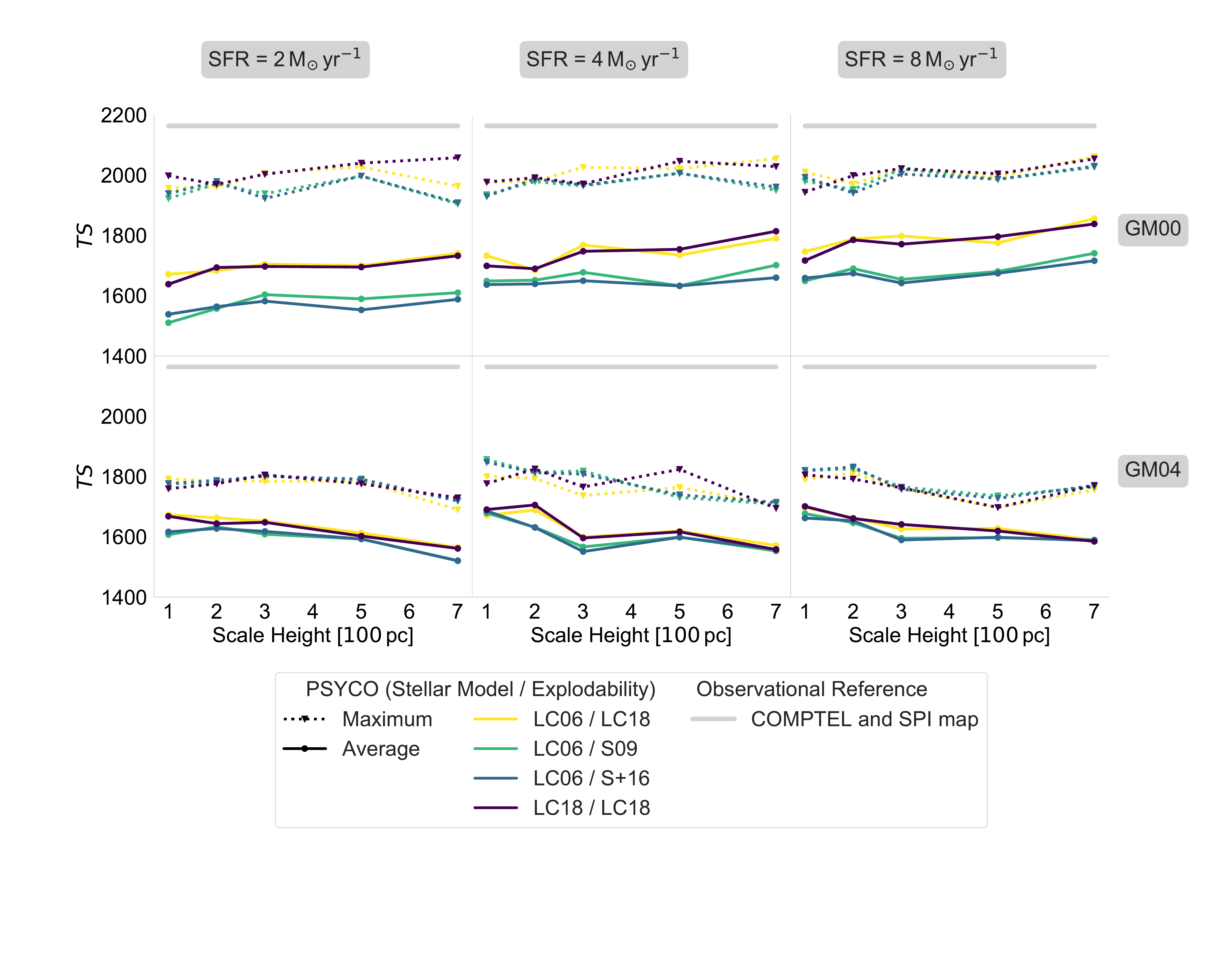}
    \caption{Likelihood ratio of 6000 sky maps modelled with PSYCO relative to the likelihood of a background-only fit with the SPI. Dots and solid lines denote the average values from 100 MC runs as a function of scale height. The colours correspond to stellar model configuration as noted in the legend. Triangles mark the maximum TS value obtained from the 100 MC samples for each model configuration. The thick gray lines denote the reference value obtained with COMPTEL ($\mrm{TS} = 2160$) and SPI ($\mrm{TS} = 2166$).}
    \label{fig:likelihood_values}
\end{figure*}

\section{Comparison to Data}\label{sec:comparison}
\subsection{Galactic \Al and \Fe fluxes}\label{sec:fluxes}
The total flux as well as the flux distribution of these models, either across the celestial sphere or along different lines of sight, play a major role in the interpretation of the \gray signals.
The total flux in \gray measurements is nearly independent of the chosen morphology (e.g., the SPI or COMPTEL maps, or an exponential disk lead to the same fluxes within 5\,\%), so that the absolute measurements of $F_{26} = (1.71 \pm 0.06) \times 10^{-3}\,$\flux \citep{Pleintinger:2019tq} and $F_{60} = (0.31 \pm 0.06) \times 10^{-3}\,$\flux \citep{Wang:2020fe} are important model constraints.
Indeed, the chosen density profile has a considerable impact on the total \Al and \Fe fluxes, changing by 50\,\%, with GM03 typically showing the lowest fluxes and GM01 the highest.
In addition to the total flux, the `Inner Galaxy' ($|l| \leq 30$\degree, $|b| \leq 10$\degree) is frequently used for comparisons in data analyses because in this range, the surface brightness of \Al (and supposedly \Fe) is particularly high.
We summarise an evaluation of \Al and \Fe fluxes in Tab.\,\ref{tab:model_fluxes} for both the entire sky and the Inner Galaxy.
This is obtained by an average across density profiles and therefore includes an intrinsic scatter of 25\,\%.

The scale height $z$ of the density profile also influences the total fluxes, with smaller scale heights typically resulting in larger fluxes for otherwise identical parameter sets.
The effect is stronger for \Fe than for \Al because a larger scale height leads to a larger average distance of sources to the observer (the galaxy is `larger').
For density profiles which show an enhancement closer to the observer (GM00, GM01, GM04), this effect is stronger than for more centrally-peaked profiles.
In addition, the later \Fe ejection compared to \Al preferentially fills older and larger bubbles.
The \Fe emission is intrinsically more diffuse so that an additional vertical spread enhances the $r^{-2}$-dependence of the flux which amplifies the scale height effect for \Fe.
We show the radial distributions of expected flux contributions for \Al and \Fe in Fig.\,\ref{fig:radial_profiles}.
Bright regions indicate `where' the most measured flux would originate in.
Clearly in all profiles, the local flux contributions, and especially the spiral arms (GM00--GM03), shape the resulting images (Fig.\,\ref{fig:example_images}).
Even though there is a density enhancement in GM00, for example, the resulting image would appear devoid of such a feature because the $r^{-2}$ effect lets the Galactic centre feature appear washed out.
Interestingly, the exponential profiles (GM03, GM04) show both, a central flux enhancement as well as a local contribution, which might be closer to the expected profile of classical novae in the Milky Way plus massive star \Al emission.
However, such a strong central enhancement is not seen in either the COMPTEL map nor the SPI map, even though exponential disks nicely fit the raw data of the two instruments.
Interpretations are discussed in Sect.\,\ref{sec:discussion}.

\begin{table}[t]
\caption{Fluxes of \Al emission at 1.809\,MeV and of \Fe emission at 1.173 or 1.332\,MeV, respectively, in units of $10^{-4}$\,\flux from PYSCO simulations for the entire sky or the Inner Galaxy ($|l| \leq 30$\degree, $|b| \leq 10$\degree), as a function of SFR (in units of \Msol\,$\mrm{yr^{-1}}$) and different stellar evolution models. The uncertainty in each value is estimated to 25\,\%, from variations of values over the different density profiles GM00--GM04.}
\begin{center}
\begin{tabular}{cc||cccc|cccc}
    \hline
    \hline
    \multicolumn{2}{c||}{Sky region} & \multicolumn{4}{c|}{Full sky} & \multicolumn{4}{c}{Inner Galaxy}\\
    \multicolumn{2}{c||}{SFR} & 1 & 2 & 4 & 8 & 1 & 2 & 4 & 8\\
    \hline
    LC06 & \Al & 1.2 & 2.6 & 5.7 & 13.0 & 0.4 & 1.0 & 2.1 & 5.0 \\
         & \Fe & 0.3 & 0.7 & 1.6 &  3.7 & 0.1 & 0.3 & 0.6 & 1.4 \\
    LC18 & \Al & 0.5 & 1.2 & 2.9 &  6.2 & 0.2 & 0.5 & 1.0 & 2.4 \\
         & \Fe & 0.5 & 1.2 & 2.3 &  5.1 & 0.2 & 0.5 & 1.0 & 2.2 \\
    
    \hline
\end{tabular}
\end{center}
\label{tab:model_fluxes}
\end{table}

\subsection{Likelihood comparisons}\label{sec:likelihood_tests}
MeV \gray telescopes cannot directly `image' the sky -- the typically shown all-sky maps are individual reconstructions (realisations) of a dataset projected back to the celestial sphere, assuming boundary conditions and an instrumental background model.
As soon as those datasets change or another instrument with another aperture and exposure is used, also the reconstructions can vary significantly.
Likewise, different realisations of \ac{PSYCO}, even with the same input parameters of density profile, scale height, \ac{SFR}, yield model, and explodability, will always look different due to the stochastic approach.
Therefore, a comparison of individual models in `image space' can -- and will most of the time -- be flawed.

In order to alleviate this problem, the comparisons should happen in the instruments' native data spaces.
This means that any type of image is to be convolved with the imaging response functions and compared in the native data space.
By taking into account an instrumental background model, then, the likelihoods of different images (all-sky maps) can be calculated and set in relation to each other.
As an absolute reference point (likelihood maximum), we use the \Al all-sky maps from COMPTEL \citep{Pluschke:2001voa} and SPI \citep{Bouchet2015_26Al}.

Clearly, individual morphological features of the Milky Way may not be mapped in all realisations of \ac{PSYCO}, which will result in `bad fits'.
These discrepancies are expected as they are mostly dominated by random effects of the particular distribution of superbubbles in the Galaxy and in the \ac{MC} simulations.
One particular realisation is therefore not expected to match all (relevant) data structures.

An almost direct comparison is nevertheless possible to some extent:
We use the INTEGRAL/SPI \Al dataset from \citet{Pleintinger:2019tq} to investigate which of the large-scale parameters are required to maximise the likelihood.
Ultimately, this results in an all-sky map (or many realisations thereof) which can be compared to SPI (and other) data.
In particular, we describe our sky maps as standalone models $M$, which are converted to the SPI instrument space by applying the coded-mask pattern for each of the individual observations in the dataset \citep[see][for details on the dataset as well as the general analysis method]{Pleintinger:2019tq}.
An optimisation of the full model, including a flexible background model \citep{Diehl2018_BGRDB,Siegert2019_SPIBG} and a scaling parameter for the image as a whole, results in a likelihood value for each model emission map.
We note that this does not necessarily provide an absolute goodness-of-fit value.
Instead, a relative measure of the fit quality can be evaluated by a test statistic,
\begin{equation}
    \mrm{TS} = 2( \log(\mathscr{L}(D|M_0)) - \log(\mathscr{L}(D|M_1)))\mrm{,}
    \label{eq:test_statistic}
\end{equation}
with model $M_1$ describing the general case of an image plus a background model, and $M_0$ describing only the instrumental background model.
The likelihood $\mathscr{L}(D|M_0)$ of the data $D$ given the background model then describes the null-hypothesis, which is tested against the alternative $\mathscr{L}(D|M_1)$.
TS values can hence be associated to the probability of occurrence by chance of a certain emission map in the SPI dataset.
The higher the TS values are, the `better fitting' an image therefore is.

We show a summary of the TS values for GM00 (best cases) and GM04 (typically used model) in Fig.\,\ref{fig:likelihood_values}.
Independent of the actual \ac{SFR}, the fit quality is almost the same, showing that the \ac{SFR} has no morphological impact between 2 and 8\,\Msolyr with GM00 resulting in a slightly higher \ac{SFR}.
This is understood because the amplitude of the emission model, i.e. one scaling parameter for the entire image, is optimised during the maximum likelihood fit.
Only models with $\mrm{SFR} \geq 4$\,\Msolyr show a fitted amplitude near 1.0 which means that SFRs of less than 4\,\Msolyr mostly underpredict the total \Al flux.
The density profiles GM03 and GM04 (exponential profiles) show on average lower TS values than Gaussian profiles, independent of the chosen spiral structure.
The reason for these apparently worse fits compared to GM00--GM02 is the central peakedness of the exponential profiles, which is absent in actual data.
The highest average TS values are found with GM02, which can be described by a protruding and rather homogeneous emission from the Inner Galaxy (Fig.\,\ref{fig:example_images}, middle), mainly originating form the nearest spiral arm.
However, high-latitude emission is barely present in GM02 because the density maximum is about 2.5\,kpc away from the observer.
In general, a flatter gradient between the two nearest spiral arms describes the data better than a steep gradient.

\begin{figure}[t]
    \centering
    \includegraphics[width=\columnwidth]{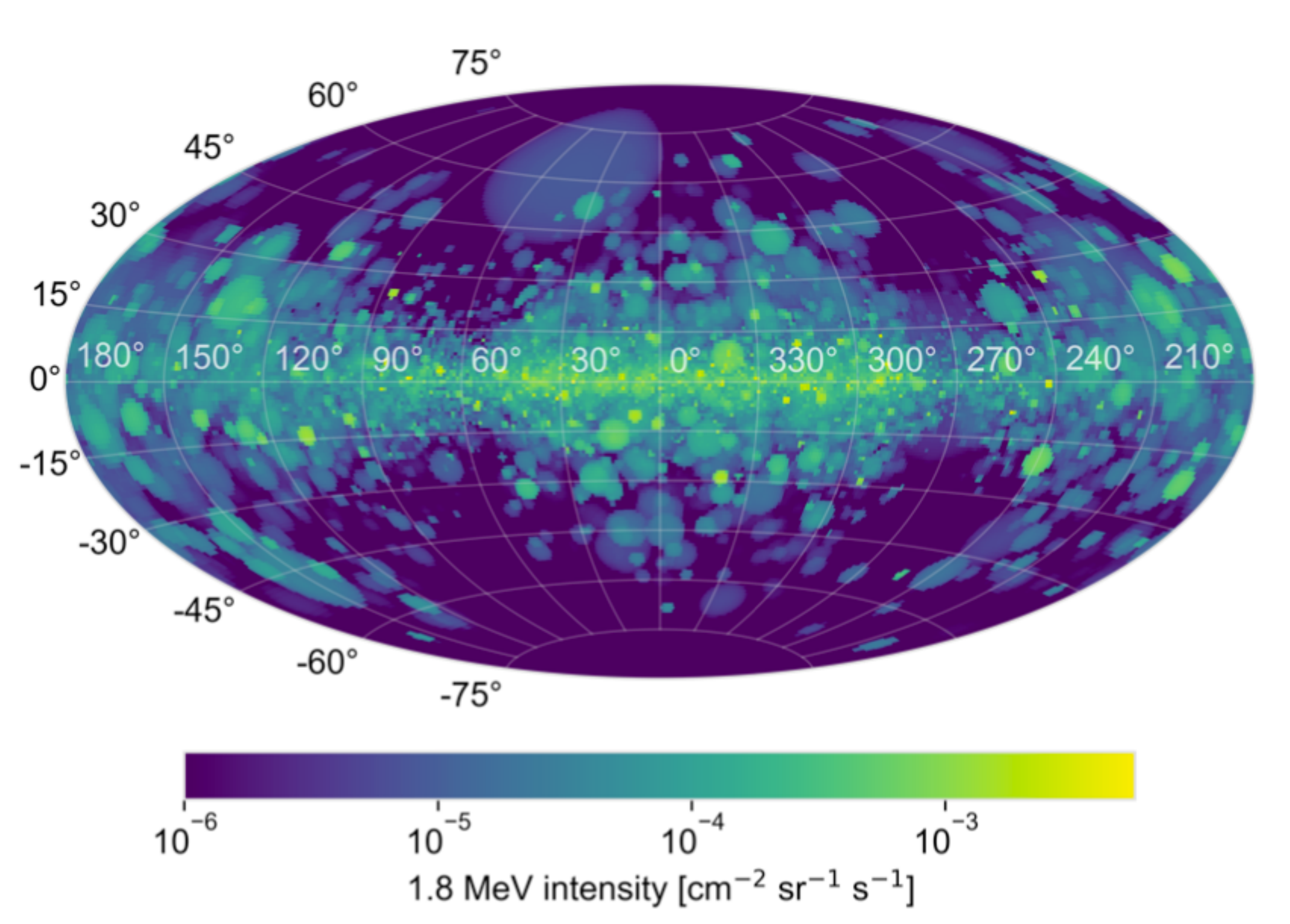}
    \caption{Best fitting sky map with $\mrm{TS} = 2061$ out of 30000 PSYCO models. It is based on GM00 with 700\,pc scale height, $\mrm{SFR} = 8$\,\Msolyr, IMF K01, stellar models LC06, and explodability LC18.}
    \label{fig:best_map}
\end{figure}

The scale height affects the quality of the global fit with a trend of a higher scale height generally fitting the data worse, except for the density profile GM00.
A large scale height increases the average distance to the supperbubbles, which decreases their apparent sizes and the general impact of foreground emission.
However, since GM00 shows the largest individual TS values (i.e. fits the SPI data best among all combinations considered), which also improves further with larger scale heights, this indicates that SPI data includes strong contributions from high latitudes in the diretion of the galactic anticentre.
In fact, GM00 is the only density profile that shows improved fits with both increasing SFR and larger scale height.
Compared to the other density profiles, GM00 requires a large SFR to explain the total flux and to fully develop its characteristics on the galactic scale because it is also the most radially extended profile reaching up to 15\,kpc.

\begin{figure}[t]
	\centering
	\includegraphics[width=\columnwidth, trim=2cm 8cm 2cm 3cm, clip]{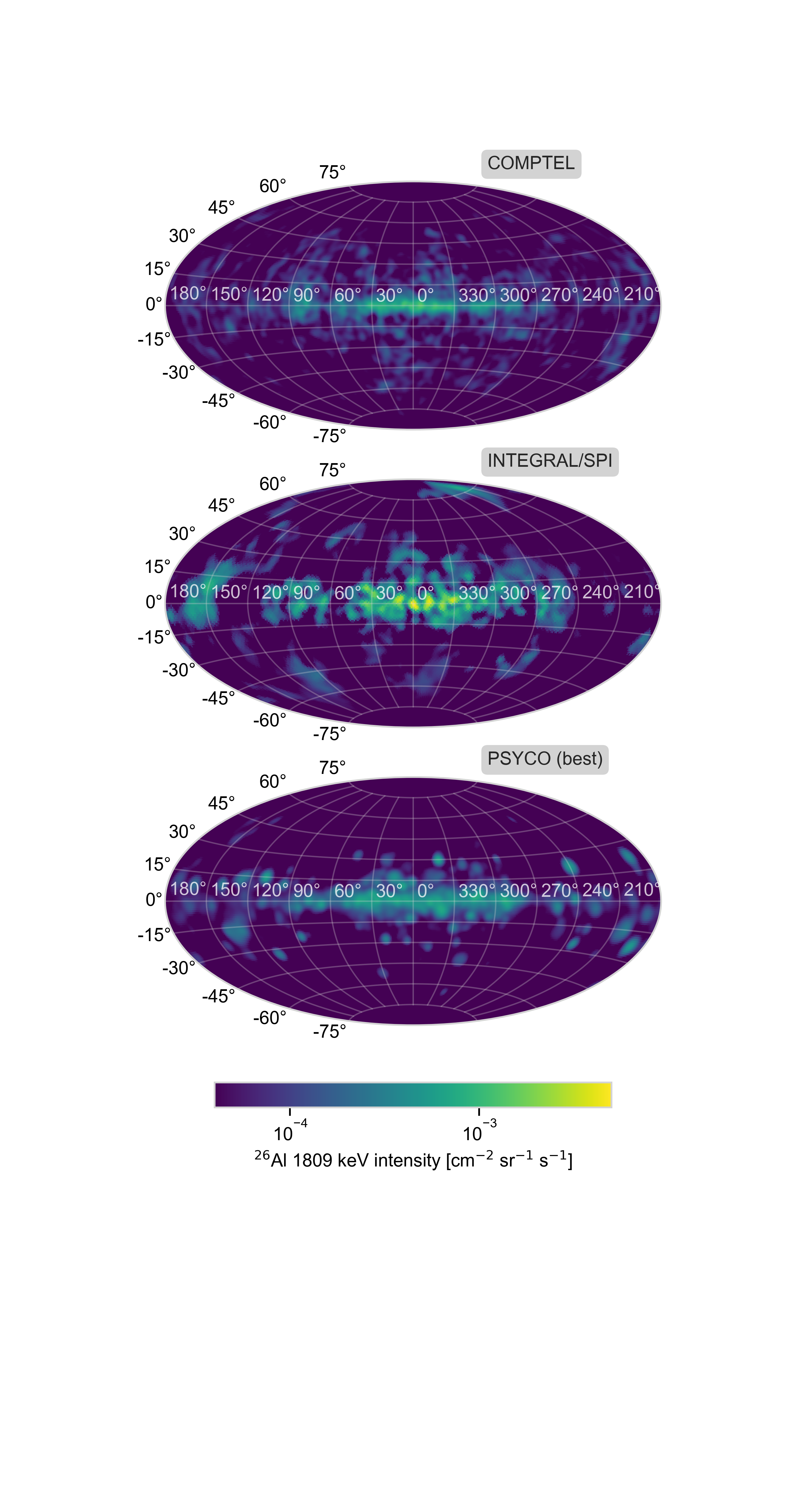}
	\caption{Compilation of observational maps (top: COMPTEL; middle: SPI) compared to our best-fitting PSYCO simulation, adopted to match the instrument resolution of 3\degree. The minimum intensity in the maps is set to $5 \times 10^{-5}\,\mrm{ph\,s^{-1}\,cm^{-2}\,sr^{-1}}$ to mimic potentially observable structures.}
	\label{fig:obsmaps}
\end{figure}

The individually best model is found for GM00 with a scale height of 700\,pc, a \ac{SFR} of 8\,\Msolyr, IMF K01, stellar evolution model LC06, and explodability L18, and shown in Fig.\,\ref{fig:best_map}.
With a TS value of 2061, it is about $\Delta\mrm{TS} = 100$ away from the maximum of the SPI map itself with ($\mrm{TS} = 2166$).
The characteristics of this map is a rather bimodal distribution, peaking toward the Inner Galaxy and the Galactic anticentre.
The spiral arm `gaps' (tangents) are clearly seen in this representation, and the map appears rather homogeneous (many bubbles overlapping to smear out hard gradients).
The map in Fig.\,\ref{fig:best_map}, however, would result in an `unfair' comparison to the reconstructed maps of COMPTEL or SPI with an intrinsic resolution of 3\degree, and a sensitivity of much more (i.e. worse) than the minimum depicted one of $10^{-6}\,\mrm{ph\,s^{-1}\,cm^{-2}\,sr^{-1}}$.
We therefore convolve the image with a 2D-Gaussian of width 2\degree, and set the minimum intensity of the maps in Fig.\,\ref{fig:obsmaps} to $5 \times 10^{-5}\,\mrm{ph\,s^{-1}\,cm^{-2}\,sr^{-1}}$.
Clearly, many features of the actual map disappear because the sensitivities of the instruments are not good enough so that especially high-latitude features beyond $|b| \gtrsim 30$\degree would drown in the background.

\section{Discussion}\label{sec:discussion}
\subsection{Degeneracy between superbubble physics, yields, and star formation}\label{sec:caveats}
The PSYCO model includes assumptions, which cannot be verified individually, yet they interplay to create the structures of the output maps.
In particular, nucleosynthesis yields and star formation rates both scale the total flux predicted by the model.
Observable structures depend on the size adopted to be filled with nucleosynthesis ejecta: we use one location centred on the massive-star group as a whole, and distribute the cumulative ejecta from the group within a spherical radius determined by the group age.
Projecting this onto the sky with the adopted distance of the group to the observer, a circular patch of \Al emission results on the map.
We know that real cavities around massive stars are not spherical, but of irregular shapes (see the example of Orion-Eridanus \citep{Brown1995_OrionOB1,Bally:2008oe}, and discussion in \citet{Krause:2013aa,Krause:2014aa}).
Also, as discussed above, the wind- and SN-blown cavities may not be filled homogeneously with \Al and \Fe, respectively.
Therefore, we caution that values of the SFR and SNR discussed in the following should be viewed as dependent in detail on the validity of these assumptions.

\subsection{Star formation and supernova rate}\label{sec:SFR_SNR}
The fluxes measured by SPI in the 1.8\,MeV line from \Al and the 1.3\,MeV line from \Fe, when fitted to the PSYCO bottom-up model, lead to SFRs of $\gtrsim 4$\,\Msolyr when considering the entire Galaxy, and $\sim 5$\,\Msolyr when considering the Inner Galaxy.
These values are obtained for the stellar evolution models by \citet{Limongi:2006gh} and \citet{Limongi:2018gh}, explodability models of \citet{Smartt:2009cc,Janka:2012sn,Sukhbold:2016aa}, and the range of velocties, metallicities and scale heights tested.
Considering uncertainties and variations with such assumptions, the SNR in the Milky Way is estimated to be 1.8--2.8 per century.
This range is purely determined by systematic uncertainties from model calculations; the statistical uncertainties would be on the order of 2--4\,\%.
Furthermore, this \ac{SFR} sets the total mass of \Al and \Fe in the Galaxy to be 1.2--2.4\,\Msol and 1--6\,\Msol, respectively.

\subsection{\Fe/\Al ratio}\label{sec:60Feto26Al}
The galactic-wide mass ratio of \Fe/\Al mainly depends on the stellar evolution model and the assigned explodabilities.
We find a mass ratio of 2.3 for model/explodability combination LC06/S09 and 7.1 for LC18/LC18.
This converts to expected flux ratios $F_{60}/F_{26}$ of 0.29 and 0.86, respectively, independent of the region in the sky chosen, i.e. either full sky or Inner Galaxy.
These values are higher compared to measured value of $0.18 \pm 0.04$ \citep{Wang:2020fe}, which is mainly due to the larger measured \Al flux compared to the PSYCO values.
Measurements of the \Fe \gray lines are currently difficult because radioactive build-up of $^{60}$Co in SPI leads to an ever-increasing background in these lines \citep{Wang:2020fe}.
Therefore, the information value of the current Galactic \Fe/\Al should not be over-interpreted.
In fact, \citet{Wang:2020fe} also estimated systematic uncertainties 
of the \Fe/\Al flux ratio to reach up to 0.4, which would be consistent with a range of parameter combinations of \ac{PSYCO}.
In fact, an increased \Fe flux from measurements might also be possible considering current detection significances of the two \Fe lines combined of $5\sigma$.
Similar to the differences in Fig.\,\ref{fig:best_map}, showing the total emission, and Fig.\,\ref{fig:obsmaps}, showing only `detectable' features, the generally higher flux of \Al throughout the Galaxy compared to \Fe might enhance the observed discrepancy even further.

The total \Al flux is consistently underestimated in PSYCO compared to measurements.
Either the total \Al mass is underestimated (also ejected per star, for example), or there is a general mismatch in the density profiles (spiral arms, foreground emission).
Only some extreme model configurations, e.g. with $\mrm{SFR} = 8$\,\Msolyr and no explodability constraints (S09), can reach the high observed fluxes.
The density profiles have a 50\,\% impact on the total Galactic flux.
The Inner Galaxy only contributes to 16\,\% on average, but can be measured more accurately with SPI because of the enhanced exposure time along the Galactic plane and bulge.
The Inner Galaxy contribution to the total sky varies in PSYCO between 23\,\% for centrally peaked profiles and 40\,\% for spiral-arm dominated profiles.
This can be interpreted as requiring an additional strong local component at high latitudes, or a particularly enhanced Local Arm toward the Galactic anticenter.
A bright spiral arm component in the Milky Way is supported by a general latitudinal asymmetry from the fourth quadrant, ($-90^{\circ} \leq l \leq 0^{\circ}$) \citep{Kretschmer:2013aa,Bouchet2015_26Al}.
For example, a more prominent Sagittarius-Carina arm could achieve such an asymmetry.

Comparing the scale sizes of \ac{PSYCO} \Al with \Fe simulations (Fig.\,\ref{fig:example_images}), it appears that scale radius and scale height should differ.
\Fe \gray emission rarely appears toward the Galactic anticentre, while it is even required in \Al emission.
\citet{Wang:2020fe} found a \Fe scale radius of $3.5^{+2.0}_{-1.5}$\,kpc and a scale height of $0.3^{+2.0}_{-0.2}$\,kpc.
Using the measured \Al values by \citet{Pleintinger:2019tq,Wang:2020fe} of $7.0^{+1.5}_{-1.0}$\,kpc and $0.8^{+0.3}_{-0.2}$\,kpc, respectively, it is clear that meaningful comparisons rely on the quality of \Fe measurements.
In our simulations, about 50\,\% of the total \Al line flux originates from within a radius between 2.8–-6.0\,kpc.
The \Fe fluxes originate from an even smaller region up to only 4\,kpc radius, with a tendency toward the Galactic centre, so that only the Inner Galaxy appears bright in \Fe emission.
If strong foreground sources are present, this fraction can even be as high as 90\,\% for \Al, however remains rather stable for \Fe.

Our density model GM00 best describes the SPI \Al data, also putting a large emphasis on the Local Arm as well as the next nearest arms.
With a large scale height of 700\,pc also nearby sources can be mimicked -- underlining the importance of foreground emission once more.

\subsection{Foreground emission and superbubble merging}\label{sec:bubble_merging_foreground}

The explodability of LC18 generally describes the 1.8\,MeV SPI data better than other combinations.
This is related to a trend to fill superbubbles with the majority of \Al later.
Thus, larger bubbles would appear brighter.
The average diameter of \Al-filled superbubbles in the Galaxy is about 300\,pc.
The underestimation in models of both the 1.8\,MeV local foreground components and the average bubble size indicates that the actual local star formation density in the vicinity of the Solar System is larger than the average, and that it occurs overall more clustered than currently assumed.
Our PSYCO full-sky images, when compared to SPI and COMPTEL, might support this as they appear more structured than in the simulations by \citet{Fujimoto:2018aa}, for example.
Clustered star formation releases energy more concentrated through stellar feedback processes.
As a result, the average size of the superbubbles would be larger and such a mechanism could also account for the more salient granularity in the observed scale height distribution of the Milky Way \citep{Pleintinger:2019tq}.
An increased bubble size might point to frequent superbubble merging, as suggested by \citet{Krause:2013aa,Krause:2015aa} or \citet{Rodgers-Lee:2019al}.
Here, HI shells break up frequently and open up into previously blown cavities, which lets them grow larger as a consequence of the feedback contributions from multiple star groups.

\subsection{Superbubble blowout and Galactic wind}\label{sec:bubble_blowout}
Using simulations similar to \citet{Rodgers-Lee:2019al}, \citet{Krause2021_26Alchimneys} have characterised the vertical blowout of superbubbles from the Galactic disc.
In some of their simulations, the superbubbles tend to merge in the disc and create transsonic, \nuc{Al}{26} carrying outflows into the halo.
With typical velocities of $1000\,\mrm{km\,s^{-1}}$ and a half life of 0.7\,Myr, kpc scale heights can be expected.
Given that massive stars are formed within the Milky Way disc within a typical scale height of about 100\,pc or less \citep{Reid2016_spiralarm_distance}, the blowout interpretation appears to be a likely explanation for the large scale heights we find. 

\citet{Rodgers-Lee:2019al} also show that the halo density constrains such vertical blowouts, and due to the higher halo density in the Galactic centre for a hydrostatic equilibrium model the scale height should be smallest there.
The apparent bimodal scale height distribution in our best-fitting models (higher towards Galactic centre and anti-centre) might hence point to a significant temporal reduction of the halo density via the Fermi bubbles \citep[e.g.,][]{Sofue2000_bipolar,Predehl2020_Xraybubbles,Yang2022_Fermierositabubbles}.

\section{Summary and outlook}\label{sec:outlook}
We have established a population synthesis based bottom-up model for the appearance of the sky in $\gamma$-ray emission from radioactive ejecta of massive star groups.
This is based on stellar-evolution models with their nucleosynthesis yields, and representations of the spatial distribution of sources in the Galaxy. 
Parameters allow to adjust these components, and thus provide a direct feedback of varying model parameters on the appearance of the sky. 
We parametrise, specifically, the explodability of massive stars, the contributions from binary evolution, the density profile and spiral-arm structure of the Galaxy, and the overall star formation rate.
PSYCO can be easily adapted to other galaxies, and e.g. model radioactivity in the Large Magellanic Cloud.

Application of the PSYCO approach to Galactic \Al finds agreement of all major features of the observed sky.
This suggests that, on the large scale, such a bottom-up model captures the sources of \Al on the large scale of the Galaxy with their ingredients. 
Yet, quantitatively the PSYCO model as based on current best knowledge fails to reproduce the all-sky $\gamma$-ray flux as observed.
This suggests that nucleosynthesis yields from current models may be underestimated.
On the other hand, mismatches in detail appear particularly at higher latitudes, and indicate that nearby sources of \Al with their specific locations play a significant role for the real appearance of the sky, and also for the total flux observed from the sky.
We know several such associations, for example Cygnus OB2 \citep{Martin2009_cygnus26al}, Scorpius Centaurus \citep{Diehl2010_ScoCen} or Orion OB1 \citep{Siegert2016_Orion26Al}, should be included for a more-realistic representation.
We note, however, that here details of superbubble cavity morphologies will be important \citep{Krause:2013aa,Krause:2015aa}, and our spherical volume approximation, while adequate for more-distant sources on average, will be inadequate.
Such refinements, and inclusions of very nearby cavities from the Scorpius-Centaurus association and possible the Local Bubble, are beyond the scope of this paper. 

Measurements of the \Al emission will accumulate with the remaining INTEGRAL mission till 2029.
The \Fe \gray lines, however, have become difficult because radioactive build-up of $^{60}$Co in SPI leads to an ever-increasing background in these lines \citep{Wang:2020fe}.
With the COSI instrument \citep{Tomsick2019_COSI} on a SMEX mission planned for launch in 2026, a one order of magnitude better sensitivity after two years could be achieved, so that, also weaker structures, as predicted from our PSYCO models could be identified in both, the \Al and \Fe lines.
Also, the Large Magellanic Cloud with an expected 1.8\,MeV flux of $2 \times 10^{-6}$\,\flux would be within reach for the COSI mission.

\begin{acknowledgements}
Thomas Siegert acknowledges support by the Bundesministerium f\"ur Wirtschaft und Energie via the Deutsches Zentrum f\"ur Luft- und Raumfahrt (DLR) under contract number 50 OX 2201.
\end{acknowledgements}

\bibliographystyle{aa} % style aa.bst
\bibliography{bibliography.bib,thomas.bib} % your references Yourfile.bib

\appendix

\section{Details on Initial Mass Functions}\label{sec:appendix_IMF}
\citep{Kroupa:2001im} refined the \ac{IMF} slope as multiple-broken power-law
\begin{equation} \label{eq:imf_kroupa}
\xi_{\text{K01}}(M_*) = k_i M_*^{-\alpha} \quad \text{with} \quad
\begin{cases}
\alpha = 0.3 & \text{for $M_* \leq 0.08$}\\
\alpha = 1.3 & \text{for $0.08 < M_* \leq 0.5$}\\
\alpha = 2.3 & \text{for $M_* > 0.5$}
\end{cases},
\end{equation}
with a normalisation constant $k_i$ in each of the three mass regimes depending on the local star formation rate.
This was adjusted by \citet{Chabrier:2003ki} to follow a smooth log-normal distribution
\begin{equation} \label{eq:imf_chabrier}
\xi_{\text{C05}}(M_*) =
\begin{cases}
\frac{a}{M_*\log(10)} \exp\left(-\frac{[\log(M_*) - \log(\mu)]^2}{2 \sigma^2}\right) & \text{if $M_* \leq 1$}\\
k_{C05} M_*^{-2.3} & \text{if $M_* > 1$}
\end{cases},
\end{equation}
with normalisation constant $k_{\text{C05}} = 1/\log(10) \exp\left[-(\log(\mu)^2/2\sigma^2)\right]$.
The log-normal parameters are amplitude $a = 0.086$, $\mu = 0.2$, and $\sigma = 0.55$ \citep{Chabrier:2005uf}.

\section{Optimal Sampling}\label{sec:appendix_optimal_sampling}
To achieve optimal sampling with a total mass $M_{\rm EC}$, the \ac{IMF} has to be normalised by excluding unphysical objects above $M_{\text{max}}$, which is determined by the relation \citep{Weidner:2004ms}
\begin{equation} \label{eq:optimal_sampling_most_massive_star}
1 = \int_{M_{\text{max}}}^{M_{\text{trunc}}} \xi(M_*)\ dM_*,
\end{equation}
with a truncation condition $M_{\text{trunc}} = \infty$.
For a complete sampling of the \ac{IMF}, it can then be discretised by the condition
\begin{equation} \label{eq:optimal_sampling_number_i}
1 = \int_{m_{i+1}}^{m_i} \xi(M)\ dM,
\end{equation}
which ensures that within each mass interval $m_i > m_{i+1}$ exactly one object is formed.
This yields the iterative formula for mass segments
\begin{equation}\label{eq:optimal_sampling_segments}
	m_{i+1} = \left(m_i^{1-\alpha} - \frac{1-\alpha}{kM_{\text{max}}^{\alpha}}\right)^{\frac{1}{1-\alpha}},
\end{equation}
with the normalisation $k$ of the \ac{IMF}.
The individual masses of each star in mass segment $i$ is then given by
\begin{equation} \label{eq:optimal_sampling_mass_i}
M_i = \frac{k}{2-\alpha} (m_i^{2-\alpha} - m_{i+1}^{2-\alpha})M_{\text{max}}^{\alpha}, \qquad \text{for}\ \alpha \neq 2.
\end{equation}
This sets the discretisation of a total cluster mass into single star masses in our population synthesis model.

\section{Line of sight integration for spherical supernubbles}\label{sec:appendix_los_integration}

For an observer at position $\boldsymbol{p}_0 = (0, 0, 0)$ with respect to the centre of the emitting sphere $\boldsymbol{s}_0 = (x_0, y_0, z_0)$, \grays can be received along each line of sight
\begin{equation}\label{eq:ray}
\boldsymbol{s}(\phi, \theta) = s \cdot \begin{pmatrix}
\cos\phi\ \cos\theta\\
\sin\phi\ \sin\theta\\
\sin\theta
\end{pmatrix}
\end{equation}
for azimuth angle $\phi$ in the Galactic plane and zenith angle $\theta$.
Only along lines of sight intersecting the sphere photons are received.
These intersections are calculated as
\begin{equation}
s_{\substack{\text{max}\\\text{min}}} = p_0 \pm \sqrt{p_0^2 - \boldsymbol{s}_0^2 + R_{\text{SB}}^2},
\end{equation} 
for the nearby and distant points from the observer $s_{\text{min}}$ and $s_{\text{max}}$, respectively.
Here, the auxiliary variables $p_0 \coloneqq x_0 \cos\phi\cos\theta + y_0 \sin\phi \cos\theta + z_0 \sin\theta$ and $\boldsymbol{s}^2_0 \coloneqq x_0^2 + y_0^2 + z_0^2$ were introduced.
For an observed outside a sphere, the differential flux across the surface of the sky is therefore
\begin{equation}
    F(\phi,\theta) := \frac{1}{4\pi} \int_{s_{\rm min}}^{s_{\rm max}}\,ds\,\rho(s) = \frac{2\rho_n}{4\pi}\sqrt{p_0^2 - \boldsymbol{s}_0^2 + R_{\text{SB}}^2}\mrm{,}
    \label{eq:los}
\end{equation}
where $\rho_n = \frac{p_{E_{\gamma}} \Msol}{\tau_n M_{n,\rm u} V}$ is the emissivity (in units of $\mrm{s^{-1}\,cm^{-3}}$) with the variables defined in Eq.\,(\ref{eq:unit_luminosity}), and $V$ being the volume contributing to the total luminosity as
\begin{equation}
    L = \int_{\rm sky}\,d\Omega\,\int_{s_{\rm min}}^{\rm s_{\rm max}}\,ds\,s^2\rho(s) = \rho_n V\mrm{.}
    \label{eq:luminosity_integral}
\end{equation}
For an observer inside a sphere, the lower integration limit $s_{\rm min}$ equals zero, and equations Eq.\,(\ref{eq:los}) and (\ref{eq:luminosity_integral}) change accordingly.
We note that the volume integral is equal to the volume of a sphere of $V = \frac{4\pi}{3}R_{\rm SB}^3$.

\section{Trends of SFR, SNR, and \Fe/\Al as a function of explodability}\label{sec:appendix_PSYCO_trends}
\begin{figure}
    \centering
    \includegraphics[width=\hsize,trim=0cm 0cm 0cm 0cm, clip=True]{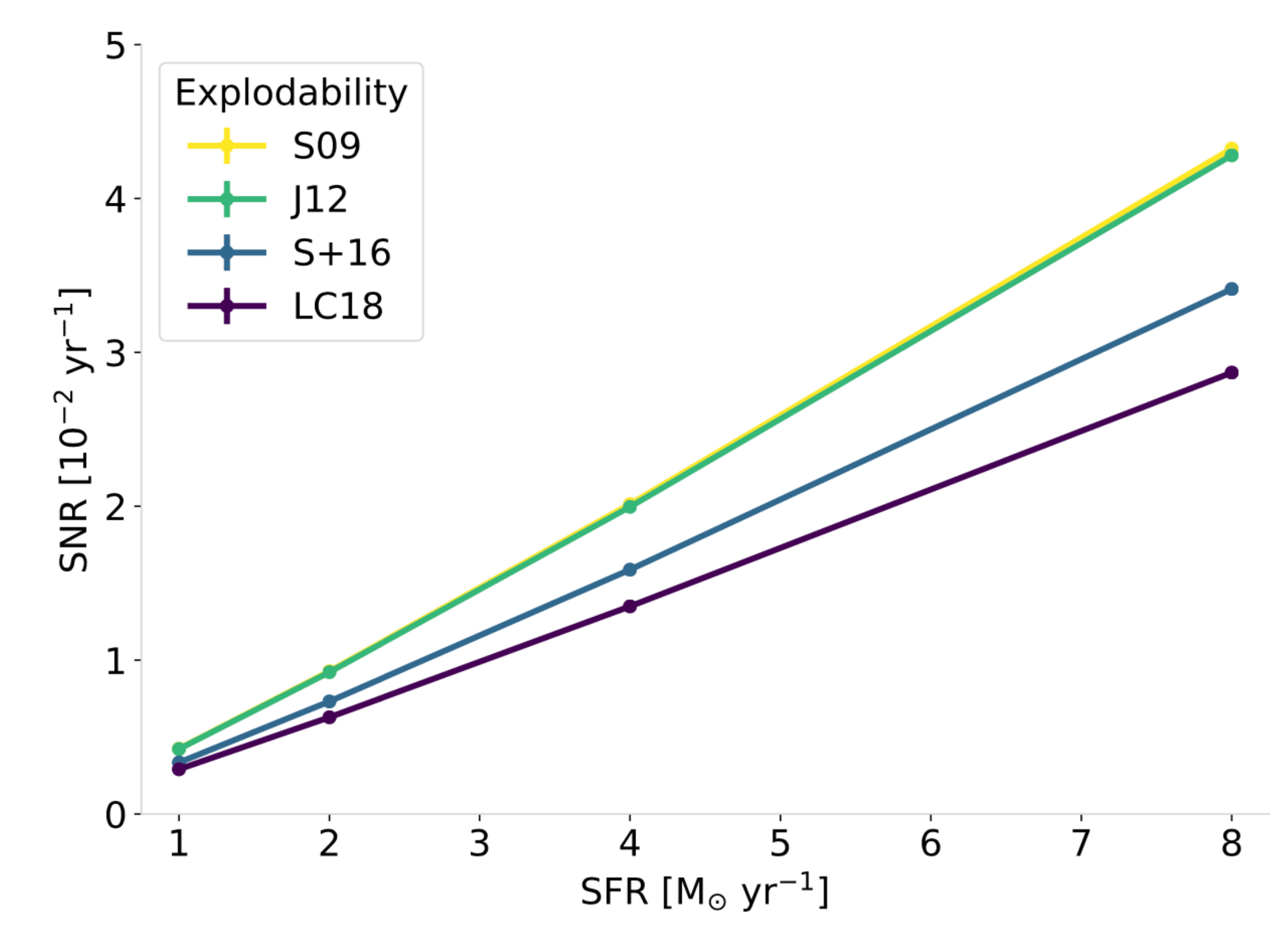}
    \caption{Supernova rate as a function of star formation rate for different explodability assumptions.}
    \label{fig:SFR_SNR_explodability}
\end{figure}

\begin{figure}
    \centering
    \includegraphics[width=\hsize,trim=0cm 0cm 0cm 0cm, clip=True]{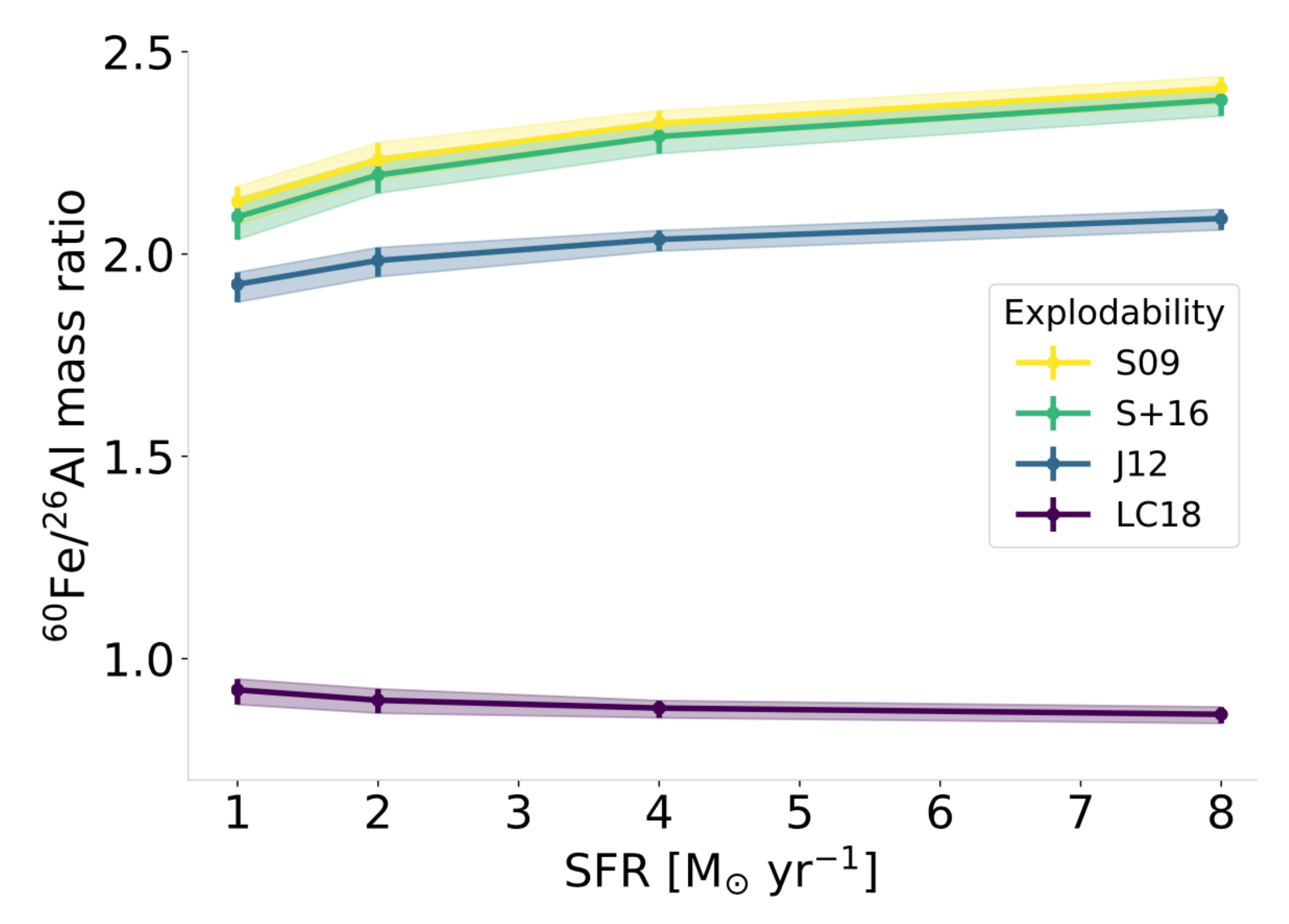}
    \caption{\Fe/\Al mass ratio as a function of star formation rate for different explodability assumptions.}
    \label{fig:Fe2Al_explodability}
\end{figure}

In Figs.\,\ref{fig:SFR_SNR_explodability} and \ref{fig:Fe2Al_explodability}, we show the effect of different explodability assumptions in the literature on the SFR, SNR, and \Fe/\Al mass ratio.
It is clear that for the rigorous assumption of no SNe above an initial stellar mass of $25\,M_{\odot}$ \citep[LC18;][]{Limongi:2018gh}, both the SNR as well as the \Fe/\Al mass ratio are decreased compared to other assumptions that include more SNe also for higher initial masses.
Especially for the measured \Fe/\Al flux ratio of $\sim 0.2$ with systematic deviations up to $\sim 0.4$ \citep{Wang2020_Fe60}, mass ratios of $1.7$ up to $3.4$ would be suggested.
However, the flux in the Galaxy is not concentrated in one point and our PSYCO simulations suggest that \Al and \Fe are not co-spatial, so that the rigorous assumption by \citet{Limongi:2018gh} may still be valid.

\end{document}